%% file: main-sample-acmsmall-conf.tex
\documentclass[acmsmall]{acmart}
\AtBeginDocument{%
  \providecommand\BibTeX{{%
    Bib\TeX}}}
\usepackage{amsmath,amsfonts}
\usepackage{graphicx}
\usepackage{textcomp}
\usepackage{tabularx}
\usepackage{caption}
\usepackage{xcolor}
\usepackage{ifthen}
\usepackage{algpseudocode}
\usepackage{multirow}
\usepackage{subcaption}
\usepackage{algorithm}
\usepackage{tcolorbox}
\usepackage{booktabs}
\usepackage{tikz}
\usepackage{soul}
\usepackage{booktabs}
\usepackage{pdflscape}
\usepackage{tikz}
\usepackage[table]{xcolor}
\usepackage[dvipsnames]{xcolor}
\usepackage[mathscr]{eucal}
\usepackage{pifont}
\newcommand{\cmark}{\ding{51}}%
\newcommand{\xmark}{\ding{55}}%

\newcolumntype{P}[1]{>{\centering\arraybackslash}p{#1}}
\DeclareMathOperator*{\argmin}{argmin} 
\DeclareMathOperator*{\argmax}{argmax} 
\def\BibTeX{{\rm B\kern-.05em{\sc i\kern-.025em b}\kern-.08em
    T\kern-.1667em\lower.7ex\hbox{E}\kern-.125emX}}

\setcopyright{cc}
\setcctype{by}
\acmDOI{10.1145/3797086}
\acmYear{2026}
\acmJournal{PACMSE}
\acmVolume{3}
\acmNumber{FSE}
\acmArticle{FSE058}
\acmMonth{7}
\received{2025-09-11}
\received[accepted]{2025-12-22}

\begin{document}

\title{Empirical Insights of Test Selection Metrics under Multiple Testing Objectives and Distribution Shifts}

\author{Jingyu Zhang}
\orcid{0000-0001-6043-4239}
\affiliation{%
  \institution{Hong Kong Metropolitan University}
  \city{Hong Kong}
  \country{Hong Kong}
}
\email{fzhang@hkmu.edu.hk}

\author{Fan Wang}
\orcid{0009-0003-5444-1813}
\affiliation{%
  \institution{City University of Hong Kong}
  \city{Hong Kong}
  \country{Hong Kong}
}
\email{fan.wang@my.cityu.edu.hk}

\author{Jacky Keung}
\orcid{0000-0002-3803-9600}
\affiliation{%
  \institution{City University of Hong Kong}
  \city{Hong Kong}
  \country{Hong Kong}
}
\email{jacky.keung@cityu.edu.hk}

\author{Yihan Liao}
\orcid{0000-0002-8002-9190}
\affiliation{%
  \institution{City University of Hong Kong}
  \city{Hong Kong}
  \country{Hong Kong}
}
\email{yihanliao4-c@my.cityu.edu.hk}

\author{Yan Xiao}
\authornote{Corresponding Author.}
\orcid{0000-0002-2563-083X}
\affiliation{%
  \institution{Shenzhen Campus of Sun Yat-sen University}
  \city{Shenzhen}
  \country{China}
}
\email{xiaoy367@mail.sysu.edu.cn}

\author{Lei Ma}
\orcid{0000-0002-8621-2420}
\affiliation{%
  \institution{University of Tokyo}
  \city{Tokyo}
  \country{Japan}
}
\affiliation{%
  \institution{University of Alberta}
  \city{Edmonton}
  \country{Canada}
}
\email{ma.lei@acm.org}

\begin{abstract}
Deep learning (DL)-based systems can exhibit unexpected behavior when exposed to out-of-distribution (OOD) scenarios, posing serious risks in safety-critical domains such as malware detection and autonomous driving. This underscores the importance of thoroughly testing such systems before deployment. To this end, researchers have proposed a wide range of test selection metrics designed to effectively select inputs. However, prior evaluations of metrics reveal three key limitations: (1) narrow testing objectives, for example, many studies assess metrics only for fault detection, leaving their effectiveness for performance estimation unclear; (2) limited coverage of OOD scenarios, with natural and label shifts are rarely considered; (3) Biased dataset selection, where most work focuses on image data while other modalities remain underexplored. Consequently, a unified benchmark that examines how these metrics perform under multiple testing objectives, diverse OOD scenarios, and different data modalities is still lacking. This leaves practitioners uncertain about which test selection metrics are most suitable for their specific objectives and contexts. To address this gap, we conduct an extensive empirical study of 15 existing metrics, evaluating them under three testing objectives (fault detection, performance estimation, and retraining guidance), five types of OOD scenarios (corrupted, adversarial, temporal, natural, and label shifts), three data modalities (image, text, and Android packages), and 13 DL models. In total, our study encompasses 1,640 experimental scenarios, offering a comprehensive evaluation and statistical analysis. 
\end{abstract}

\ccsdesc[500]{Software and its engineering~Software testing and debugging}

\keywords{Deep Learning Testing, Test Selection Metrics, Empirical Study}
\maketitle
\input{introduction.tex}

\input{background-related-work}

\input{Methodology}
\input{Experimental_setup}

\input{Results}

\input{discussion}
\input{Conclusion}

\bibliographystyle{ACM-Reference-Format}
\bibliography{sample-base}
\end{document}

%% file: introduction.tex
\section{Introduction} \label{sec: introduction}
Nowadays, Deep Neural Networks (DNNs) have been integrated into many software applications due to their great success in performance and efficiency, including safety-critical applications like autonomous driving systems. However, recent research has shown that DNNs with high accuracy on the in-distribution (ID) samples may demonstrate misbehavior on out-of-distribution (OOD) scenarios, such as corrupted \cite{kim2019guiding, hu2022empirical} and adversarial samples \cite{goodfellow2014explaining, pei2017deepxplore}. The fact that deployed DNNs often encounter OOD inputs poses severe threats to user safety and pinpoints the importance of comprehensive testing of DNNs before deployment. A common DNN testing process is to select a small representative set of inputs (often regarded as \textit{test suite}) that achieves the testing objective (e.g., detecting DNN faults) \cite{hu2024test, shi2021empirical, ma2021test}. Three main testing objectives are studied in the field of DNN testing \cite{hu2024test}: Objective 1. fault detection, which selects test samples to expose as many faults as possible. We denote a \textit{fault} in DNNs as the root reason that causes mispredictions \cite{attaoui2023black, aghababaeyan2023black}; Objective 2. performance estimation, which selects and labels samples with the goal of providing an accurate estimation of the model performance (e.g., accuracy) in the unlabeled operational set; Objective 3. retraining guidance, which selects and labels the samples that can improve the model performance as much as possible. Given that the labeling process requires extensive expert labor, the main challenge in DNN testing is to achieve the desired objective with a very limited labeling budget. To address this, a great number of test selection metrics has been proposed, where we roughly categorize them into six types based on their characteristics: uncertainty-based \cite{wang2014new,feng2020deepgini}, diversity-based \cite{pei2017deepxplore,ma2018deepgauge, aghababaeyan2023black}, surprise-based \cite{kim2019guiding}, sampling-based \cite{li2019boosting, guerriero2021operation}, clustering-based \cite{shen2020multiple, chen2020practical}, and hybrid \cite{hu2022empirical,zhou2020cost} metrics. They aim to efficiently select and prioritize a small number of test inputs. The selected inputs are then manually labeled and are expected to achieve the targeted objective. For example, DeepGini \cite{feng2020deepgini} selects uncertain inputs based on the softmax score, with the intuition that uncertain inputs can be fault-revealing and therefore enhance the model by retraining. As for the performance estimation objective, metrics like Cross Entropy-based Sampling \cite{li2019boosting} and DeepReduce \cite{zhou2020cost} are proposed to use different techniques to maximize the distribution similarity between the selected test suite and the operational set. However, we observe three key limitations in the existing work:

\textbf{(1) Narrow testing objectives studied:} 
It remains unclear whether metrics are effective beyond the objectives they were originally designed for. Among the 15 metrics \cite{feng2020deepgini, wang2014new, pei2017deepxplore, ma2018deepgauge,aghababaeyan2023black,kim2019guiding,li2019boosting,chen2020practical,zhou2020cost,guerriero2021operation,shen2020multiple,hu2022empirical} studied in this paper, 9 focus on a single objective, 5 consider two objectives, while none have been evaluated across all three. Yet practitioners may need metrics that serve multiple objectives simultaneously. For example, while DeepGini \cite{feng2020deepgini} is known to be effective for fault detection, can it also be used for performance estimation? Similarly, could metrics not designed for fault detection perform even better?

\textbf{(2) Limited coverage of OOD scenarios:} Evaluating metrics under diverse testing environments is essential to assess their generalizability. However, our preliminary survey of 26 test selection papers shows that 23 \cite{chen2020practical,li2019boosting,guerriero2024deepsample,guerriero2021operation,weiss2022simple,attaoui2023black,feng2020deepgini,kim2020reducing,pei2017deepxplore,li2024prioritizing,kim2019guiding,liang2018redefining,li2024test,aghababaeyan2024deepgd,ma2021test,byun2019input,mosin2022comparing,hu2025assessing,wang2024can,zhao2022can,shi2021empirical, harel2020neuron} of them focus primarily on original test data, 8  \cite{chen2020practical,weiss2022simple,hu2022empirical,pei2017deepxplore,li2024prioritizing,kim2019guiding,li2024test,shen2020multiple} on corrupted shifts, 11 \cite{chen2020practical, hu2022empirical,feng2020deepgini,pei2017deepxplore,kim2019guiding,shen2020multiple,hao2023test,ma2021test,zhao2022can, shi2021empirical, harel2020neuron} on adversarial shifts, only 2  \cite{hu2022empirical, wang2024can} on natural shifts and none for label shifts.  

\textbf{(3) Biased dataset selection:} 
To ensure generalizability, the effectiveness of test selection metrics should be evaluated across diverse data modalities. However, among 26 surveyed papers, image data dominate the evaluation (23/26 papers \cite{chen2020practical,zhou2020cost,li2019boosting,guerriero2024deepsample,guerriero2021operation,weiss2022simple,hu2022empirical,attaoui2023black,feng2020deepgini,pei2017deepxplore,kim2019guiding,shen2020multiple,hao2023test,aghababaeyan2024deepgd,ma2021test,byun2019input,mosin2022comparing,kim2020reducing,hu2025assessing,wang2024can,zhao2022can,shi2021empirical,harel2020neuron}), while text data (2/26 \cite{weiss2022simple,hu2022empirical}) and Android applications (1/26 \cite{pei2017deepxplore}) remain critically underexplored. 

To address these shortcomings, we conduct \textbf{a unified and extensive empirical study} that (1) compares 15 existing metrics across three testing objectives, five types of OOD scenarios (corruption, adversarial, temporal, natural, and label shifts), and three data modalities (image, text, and Android applications). To ensure fairness, we also identify the evaluation criteria that reliably reflect each testing objective. \textbf{Findings:} Metrics that encourage test input diversity are effective in objective 2, outperforming those specifically designed for this objective. Metrics designed for Objective 3 are also outperformed by other metrics. Moreover, there is no clear change in metrics' performance under different OOD types. (2) Our study also analyzes the metrics' performance under different selection budgets. \textbf{Findings:} metrics' performance on objectives 1 and 2 is largely stable across budgets, while there is considerable fluctuation for objective 3. (3) We examine the time efficiency of all 15 metrics for practical usage. \textbf{Findings:} A clear performance-speed trade-off exists for best-performing metrics in objective 1. No such trade-off is observed in objectives 2 and 3.

Our study makes the following two main contributions.

(1) To the best of our knowledge, we undertake the first extensive study aimed at systematically assessing the performance of 15 test selection metrics on three testing objectives under five OOD types with three data modalities and 13 DNNs. Our study encompasses a total of 1,640 experimental scenarios, providing a comprehensive evaluation
and statistical analysis.

(2) We present findings and provide implications for researchers and practitioners in DNN testing. For example, we recommend utilizing diversity-based metrics for an accurate performance estimation.

%% file: background-related-work.tex
\section{Background and Related Work} \label{sec: background and related work}
\subsection{DNN Testing}

To guarantee the robustness and reliability of DNNs, it is essential to conduct comprehensive testing to ensure their performance when facing data with diverse distributions (e.g., OOD scenarios) \cite{hu2025assessing}. While DNNs can demonstrate good performance on data collected from the ID scenarios, they have demonstrated severe misbehavior when encountering OOD inputs (e.g., adversarial inputs \cite{goodfellow2014explaining}) that are highly likely to occur during deployment. A common way to conduct DNN testing is to prepare representative and diverse test suites, label test inputs, and evaluate the model on these curated sets \cite{hu2024test}. The diversity of test suites can stem from data modalities, OOD scenarios, and task types. In this paper, we have explored three modalities: image \cite{lecun1998mnist, udacity_dataset}, text \cite{maas2011learning}, and Android packages \cite{allix2016androzoo}, together with five types OOD shifts \cite{moreno2012unifying} including corrupted, adversarial, natural, temporal, and label shifts, under two task types, classification and regression.

\subsection{Test Selection} \label{sec:bg-test-selection}
To reduce the manual labeling cost of large test suites \cite{hu2024test}, various test selection metrics have been proposed to choose a small yet effective test suite under a budget. In this paper, we study 15 metrics and categorize them into six types: (1) \textbf{Uncertainty-based metrics} select inputs where the model is most uncertain, Entropy (Ent) \cite{wang2014new} and DeepGini (Gini) \cite{feng2020deepgini} employ different equations to compute uncertainty scores from the softmax probability. (2) \textbf{Diversity-based metrics}: Neuron Coverage (NC) \cite{pei2017deepxplore} and K-multisection Neuron Coverage (KMNC) \cite{ma2018deepgauge} measure the diversity of a test suite through the coverage of neurons. Geometric Diversity (GD) and Standard Deviation (STD) \cite{aghababaeyan2023black} directly measure the input feature diversity using pre-trained feature extractors. (3) \textbf{Surprise-based metrics}: Likelihood-based surprise adequacy (LSA) and distance-based surprise adequacy (DSA) \cite{kim2019guiding} use kernel density estimation and Euclidean distance, respectively, to quantify the surprise of an input compared to the training data. (4) \textbf{Sampling-based metrics} utilize different sampling strategies to match the distribution of the testing set: Cross Entropy-based Sampling (CES) \cite{li2019boosting} minimizes the cross entropy between the selected set and the whole testing set. DeepEST (EST) \cite{guerriero2021operation} utilizes adaptive sampling. We also studied random sampling (Rand) as a baseline metric. (5) \textbf{Clustering-based metrics} partition tests into groups and select representative ones from each group (Multiple-Boundary Clustering and Prioritization (MCP) \cite{shen2020multiple} and Practical Accuracy Estimation (PACE) \cite{chen2020practical}). (6) \textbf{Hybrid metrics} employ multiple strategies/phases: DeepReduce (DR) \cite{zhou2020cost} selects the suite to guarantee the same NC and further optimizes by maximizing distribution similarity. Distribution-aware test selection (DAT) \cite{hu2022empirical} selects uncertain ID inputs (using Gini) and randomly selects OOD inputs.

\subsection{Test Optimization} 
These metrics are demonstrated effective in different testing objectives. There are three important objectives in test optimization \cite{hu2024test}: (1) fault detection \cite{aghababaeyan2023black, feng2020deepgini, gao2022adaptive, kim2019guiding}, (2) performance estimation \cite{chen2020practical, li2019boosting, zhou2020cost, guerriero2021operation}, and (3) retraining guidance \cite{hu2022empirical, shen2020multiple}.  

For (1), the metrics are expected to select inputs that can reveal as many faults as possible. For example, uncertainty \cite{feng2020deepgini, wang2014new}, diversity \cite{pei2017deepxplore,ma2018deepgauge,aghababaeyan2023black}, and surprise-based metrics \cite{kim2019guiding} are demonstrated to be effective in fault detection. For (2), the metrics (e.g., sampling-based \cite{li2019boosting, guerriero2021operation}, clustering-based \cite{chen2020practical}) are expected to select a small set of test inputs that can precisely estimate the accuracy of the whole unlabeled testing set. For (3), the metrics (e.g., MCP \cite{shen2020multiple}, DAT \cite{hu2022empirical}) are expected to select inputs that enhance DNN as much as possible through retraining.

\subsection{Empirical Study on DNN Testing}
Several works conduct empirical studies to investigate the effectiveness of existing metrics \cite{hu2025assessing, hu2022empirical, shi2021empirical, ma2021test, weiss2022simple}. For example, Hu \textit{et~al.} \cite{hu2022empirical} studies how metrics perform in retraining guidance when facing three types of OOD shifts. Their later work \cite{hu2025assessing} challenges metrics targeting fault detection and performance estimation using new ID test data (e.g., correctly classified but uncertain data) on image classification datasets. Ma \textit{et~al.} \cite{ma2021test} explores metrics targeted at fault detection on image data with adversarial shifts only. Moreover, Sun \textit{et~al.} \cite{sun2023robust} assess the effectiveness of metrics in fault detection and retraining guidance only on image classification with limited OOD scenarios. Similarly, Demir \textit{et~al.} \cite{demir2024test} evaluate only the fault-revealing capability of uncertainty-based metrics under original and synthetically generated OOD data for image classification. In addition to test selection, Berend \textit{et al.} \cite{berend2020cats} analyze how the test generators affect the data distribution, and do not study the multi-objective effectiveness of test selection.  
We note that none of these works have provided a unified evaluation benchmark of existing metrics under all three testing objectives, while also covering diverse data modalities and OOD types, which highlights the significance of our study.

%% file: Methodology.tex
\section{Study Design} \label{sec: study design}

This section first outlines the notations and provides a general overview of our study. We then formally define three testing objectives and their associated evaluation criteria. Finally, we introduce the types of OOD data used to construct diverse testing environments and present our research questions.

\subsection{Notations}
\noindent $\bullet \hspace{2mm} \mathcal{X}$ and $\mathcal{Y}$ denotes the input space and the output space, respectively;\\
$\bullet \hspace{2mm} M : \mathcal{X} \rightarrow \mathcal{Y}$ denotes the DNN model under test; \\
$\bullet \hspace{2mm} (x,y) \in \mathcal{X} \times \mathcal{Y}$ denotes a single input sample $x$ and its ground truth label $y$; \\
$\bullet \hspace{2mm} (X_{test},Y_{test}) \subseteq \mathcal{X} \times \mathcal{Y}$ denotes the testing dataset with samples $X_{test}$ and the corresponding set of ground truth labels $Y_{test}$. Note that it may contain both ID and OOD data samples;\\
$\bullet \hspace{2mm} f(M,X,Y)$ denotes a function measuring the performance (accuracy in our context) of $M$ when predicting $Y$ from $X$; \\
$\bullet \hspace{2mm}  (X_s, Y_s)  \subset (X_{test},Y_{test})$ denotes the data samples $X_s$ and corresponding labels $Y_s$ in the test suite selected by the test selection metric;\\
$\bullet \hspace{2mm}  (X_t, Y_t)  = (X_{test} \backslash X_s,Y_{test} \backslash Y_s)$ denotes that $X_t$ is the set of all inputs in $X_{test}$ that are not in $X_s$, with $Y_t$ as the corresponding label set;\\
$\bullet \hspace{2mm}   P(x|y)$ denotes the class-conditional distribution of seeing input $x$ given that the label is $y$; \\
$\bullet \hspace{2mm}  P(y)$ denotes the distribution of labels $y$;\\
$\bullet \hspace{2mm}  P(y|x)$ denotes the probability of a label $y$ given an input $x$;

\begin{figure*}[!htbp]
    \centering
    \includegraphics[width=0.9\linewidth]{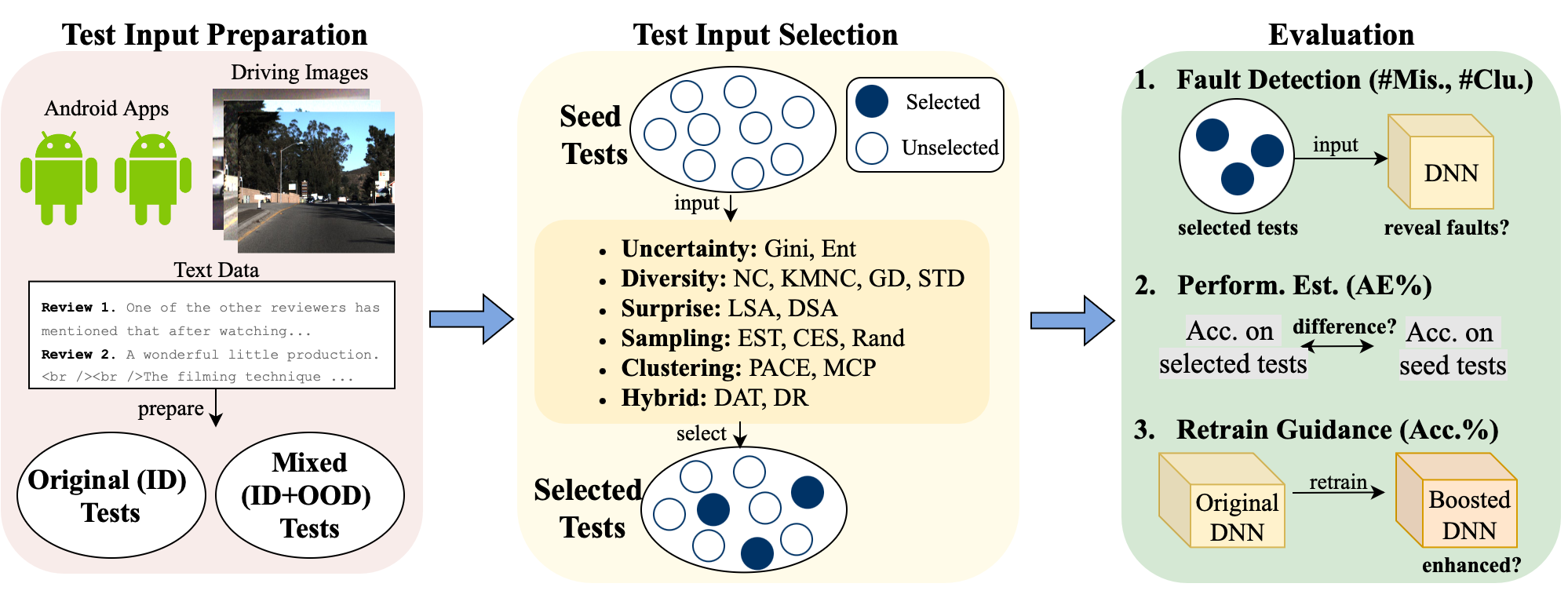}
    \caption{Overview of our study: 1. Test Input Preparation, which prepares a testing set that contains either original (ID) or mixed (ID+OOD) test samples. 2. Test Input Selection, which uses the 15 studied metrics to select inputs. 3. Evaluation: This part evaluates the selected inputs with three testing objectives: fault detection (\#Mis. and \#Clu.), performance estimation (assessed by AE\%), and retraining guidance (Acc.\%).}
    \label{fig:overview}
\end{figure*}

\subsection{Overview}
Figure \ref{fig:overview} outlines the overview, containing three main parts: test input preparation, test input selection, and evaluation. In \textbf{test input preparation}, to assess the generalized performance of test selection metrics, we design the testing sets (containing seed inputs for selection) across diverse data modalities and distribution shifts. Our benchmark evaluates \textit{three data modalities}: Android applications (AndroZoo \cite{allix2016androzoo}), images (MNIST \cite{lecun1998mnist}, Udacity \cite{udacity_dataset}), and text data (IMDb \cite{maas2011learning}). This mitigates the significant dataset selection bias observed in prior studies, where image data dominate (23 out of 26 surveyed papers), while text data (2 papers) and Android applications (1 paper) remain critically underexplored (see Section \ref{sec: introduction} for details of the survey). Furthermore, we extend our evaluation beyond original testing sets to include \textit{five OOD types}: corruption, adversarial, temporal, natural, and label shifts. This addresses another gap in the literature, as prior work has focused predominantly on original (23/26), corrupted (8/26), and adversarial (11/26) shifts, with natural shifts rarely studied (2/26) and label shifts (0/26) entirely unexplored. The inclusion of these overlooked shifts enhances the comprehensiveness and practical relevance of our evaluation benchmark. For \textbf{test input selection}, we employ 15 widely-studied metrics to select inputs from the curated testing sets. These metrics, proposed between \textit{2014 and 2023}, are categorized into \textit{six types} based on their characteristics (see Section \ref{sec:bg-test-selection} for details). Each selected test suite is then \textbf{evaluated} on \textit{three testing objectives}: fault detection (9/15 metrics have studied this objective), measured by \#Mis. and \#Clu., performance estimation (4/15), measured by AE\%, and retraining guidance (6/15), measured by Acc.\%. This comprehensive threefold analysis provides a novel comparison, as prior work has not evaluated metrics across all these objectives.

\subsection{Testing Objectives}

We provide details of testing objectives we would like to achieve with the selected test suite, and their corresponding evaluation criteria, to accurately assess the metrics' performance. 

\subsubsection{Fault Detection} \label{sec:study-design-fault-detection}
We denote a \textit{fault} in DNNs as the cause of mispredictions, such that multiple mispredictions may be attributed to the same fault. Formally, given an unlabeled testing dataset $X_{test}$, a DNN model $M$, and a labeling budget $bg$, fault detection objective aims at selecting a subset $X_s$ from $X_{test}$ such that $|X_s|=bg$ and $X_s = \argmin_{X_i \subset X_{test},|X_i|=bg} f(M,X_i, Y_i)$. $Y_i$ denotes the labels corresponding to $X_i$. We use two criteria to evaluate this objective: the number of mispredictions \cite{guerriero2024deepsample, guerriero2021operation,weiss2022simple, feng2020deepgini} and the number of clusters \cite{aghababaeyan2023black, attaoui2024supporting}.
  
\textbf{Number of Mispredictions (\#Mis.):} We measure the number of mispredictions made by the DNN under test on the selected test suite. We treat classification and regression differently. For classification tasks, we count the number of samples whose predicted class is different from their labeled class. For regression tasks, we set a threshold and count the number of samples whose offset (i.e., difference between the predicted value and the ground truth) is greater than the threshold $\delta$. Specifically, for the Udacity dataset (steering angle prediction task), we evaluate $\delta \in [0^{\circ}, 2.5^{\circ}, ..., 22.5^{\circ}, 25^{\circ}]$ and report the average number of mispredictions over these thresholds. Note that \#Mis. may not always reflect unique DNN faults when different mispredictions are attributed to the same underlying faults \cite{aghababaeyan2023black, attaoui2024supporting}

\textbf{Number of Clusters (\#Clu.):} To mitigate the limitation of \#Mis., we also adopt a clustering-based fault estimation approach \cite{aghababaeyan2023black, attaoui2024supporting} to evaluate the selected test suite, where clustering is performed on the mispredicted samples. Each cluster of mispredicted inputs in the feature space is treated as a distinct DNN fault, i.e., mispredicted inputs within the same cluster are assumed to be mispredicted for the same underlying reason. This approach (the clustering pipeline) contains three steps: feature extraction (FE), dimensionality reduction (DR), and clustering algorithm (CA). The effectiveness of this criterion depends on clustering quality, where we examine the cluster-to-fault correspondence in Section \ref{sec:RQ1.1.cluster} and identify the optimal pipeline.

\subsubsection{Performance Estimation}
Given an unlabeled testing set $X_{test}$ (also known as operational set in the context of operational testing) which possibly given as inputs to the target DNN $M$ and a labeling budget $bg$, the performance estimation objective aims at selecting a subset $X_s$ from $X_{test}$ and obtain corresponding labels $Y_s$ such that $|X_s| = bg$ and $X_s = \argmin_{X_i \subseteq X_{test}, |X_i|=bg} |f(M,X_i,Y_i)-f(M,X_{test},Y_{test})|$. In other words, the small labeled test suite is expected to accurately estimate the performance of DNN (e.g., accuracy) on the large unlabeled testing set. 

\textbf{Absolute Error (AE\%)} is a commonly used criterion to measure whether the suite selected by the metrics achieves the performance estimation objective \cite{li2019boosting,chen2020practical}, which is evaluated by $AE = |\hat{acc_i} - acc|$. $\hat{acc_i}$ and $acc$ refer to the estimated and actual accuracy (in percentage), respectively. In the implementation, $acc$ is computed from the whole testing set, whereas $\hat{acc_i}$ is calculated from the small selected set. For regression tasks, we use 1-RMSE as the accuracy. RMSE=$\sqrt{\frac{1}{N} \sum_{i=1}^{N} (pred_i - gt_i)^2}$, where $pred_i$ and denote the predicted value and the ground truth label for input $i$, respectively. $N$ denotes the total number of inputs.

\subsubsection{Retraining Guidance}

Given a pre-trained DNN $M$ with parameter weights trained on the training set $(X_{train}, Y_{train})$. Retraining guidance objective aims to select a subset $X_s$ from the unlabeled set $X_{test}$ and obtain the corresponding label $Y_s$, where $|X_s| = bg$ and $X_s = \argmax_{X_i \subseteq X_{test}, |X_i|=bg} \allowbreak f(M', X_t, Y_t)$. The improved DNN $M'$ is usually retrained on either $(X_{train} \cup X_s, Y_{train} \cup Y_s)$ (denoted as type II retraining) or $(X_s, Y_s)$ (type I retraining). As illustrated in the evaluation part (green) in Figure \ref{fig:overview}, the goal is to maximize the accuracy improvement of $M'$ (boosted DNN in orange) over $M$ (original DNN in yellow).

\textbf{Accuracy Improvement (Acc.\%)} is used to measure the capability of retraining guidance \cite{hu2022empirical,weiss2022simple,kim2019guiding}. To avoid data leakage \cite{kaufman2012leakage}, the set $(X_t, Y_t)$ used for evaluating the retrained model $M'$ does not overlap with the set $(X_s, Y_s)$ that is used in retraining. Formally, $acc' = acc(M', X_t, Y_t) - acc(M,X_t,Y_t)$. $acc(M,X_t,Y_t)$ denotes the accuracy (in percentage) of $M$ on the set $(X_t, Y_t)$.

\subsection{OOD Data Construction} \label{sec:OOD data construction}

In the real world, there are diverse types of OOD scenarios that DNNs may encounter in deployment, and it is important to understand the effectiveness of these selection metrics on different types of OOD scenarios. In this paper, we formally categorize covariate and label distribution shifts \cite{moreno2012unifying}, covering corrupted, adversarial, natural, temporal, and label shifts in total.  

\subsubsection{Covariate shift}
It refers to the changes in the input feature distribution, while the label distribution given the input is unchanged. This can be demonstrated by the change in lighting, blur, or noise for images \cite{pei2017deepxplore, mu2019mnist}. For malware data, this is commonly observed in data collected at different times, known as temporal covariate shift \cite{grosse2017adversarial, mclaughlin2017deep}. 

\textit{ \textbf{Definition.} Covariate shift is defined as the case where $P_{id}(y|x)=P_{ood}(y|x)$ and $P_{id}(x) \neq P_{ood}(x)$.}

For MNIST, IMDb, and Udacity datasets, we use three types of covariate shifts: corrupted, adversarial, and natural shifts \cite{serban2018adversarial}. For AndroZoo (data collected in 2017 \cite{allix2016androzoo}), we use temporal (data in 2018 and 2019 \cite{allix2016androzoo}), adversarial, and natural shifts. To simulate a generalized adversarial scenario, for MNIST, Udacity, and AndroZoo, we use a combination of adversarial images generated by FGSM \cite{goodfellow2014explaining}, BIM \cite{kurakin2018adversarial}, and PGD \cite{madry2017towards}, on three equal parts, respectively. For IMDb, we use Probability Weighted Word Saliency (PWWS) \cite{ren2019generating} and Banana Word Swap \cite{bananawordswap} to generate a natural and semantically valid adversarial text. For corrupted OOD shifts, we publicly available corrupted datasets from \cite{mu2019mnist} and \cite{weiss2022simple} for MNIST and IMDb, respectively. To the best of our knowledge, there is no corrupted version of the Udacity dataset in the literature. Therefore, we created the corrupted dataset, where images are corrupted using all methods from the Python package \texttt{imagecorruptions} \cite{michaelis2019dragon}, which contains 15 common visual degradations (e.g., blur, noise, weather effects) for real-world image data. For natural covariate shift, we use EMNIST \cite{cohen_afshar_tapson_schaik_2017}, Dave \cite{Dave_testing_dataset}, Drebin \cite{arp2014drebin}, and Customer Review \cite{hu2004mining} for MNIST, Udacity, AndroZoo, and IMDb, respectively. 

\subsubsection{Label shift}
The label shift refers to changes in the distribution of the label variable $y$, but the feature distributions conditional on the labels are fixed \cite{chen2022estimating}.

\textit{ \textbf{Definition.} It is defined as the case where $P_{id}(x|y) = P_{ood}(x|y)$ and $P_{id}(y) \neq P_{ood}(y)$.}

The goal is to change the label distribution in the testing set $P_{ood}(y)$ away from the training distribution $P_{id}(y)$. For MNIST, we simulate label shifts by sampling a skewed distribution, i.e., we choose roughly 34\% for digit 0, 15\% each for digits 1-3, 5\% each for digits 4-6, 2\% each for digits 7-9. For IMDb, we simulate sentiment skew by sampling 80\% positive and 20\% negative reviews, reflecting real-world user feedback distributions. For the AndroZoo dataset, malware examples are much fewer than goodware. We simulate the label shift by oversampling malware samples and undersampling goodware samples (we choose 80\% malware and 20\% goodware) to create a realistic high-threat environment. For the Udacity dataset, steering angles are predominantly centered around 0 (straight driving). To simulate label shift, we emphasize sharp turns by reducing the frequency of near-zero steering angles and increasing the presence of extreme left and right turns. Specifically, we set 40\% of samples with $y<-0.2$, 40\% with $y>0.2$ and only 20\% represent near-zero angles ($|y|\leq 0.2$).

\subsection{Research Questions}
\textit{RQ1.1. Selection of optimal clustering pipelines.}
It is crucial to validate whether each cluster can reliably represent a DNN fault. This research question investigates (1) the clustering quality and (2) the cluster-to-fault correspondence. 
For (1), we study several candidate techniques used in the three-step clustering pipeline (described in Section \ref{sec:study-design-fault-detection}) and leverage Silhouette \cite{rousseeuw1987silhouettes} and DBCV \cite{moulavi2014density} scores to assess the clustering quality of each candidate pipeline and identify the best one for each dataset. For (2), we conduct feature pattern inspection and cluster-specific retraining for the best pipeline to validate whether clusters can meaningfully represent different faults.

\textit{RQ1.2. Selection of optimal retraining processes.}
 To evaluate retraining guidance fairly, a uniform and effective retraining process ensures comparability across metrics and reveals the true margin of improvement achievable by selected test suites. In this research question, we compare type I (retrain on $(X_s, Y_s)$) and type II (retrain on $(X_{train} \cup X_s, Y_{train} \cup Y_s)$) retraining processes.

\textit{RQ2. Performance of test selection metrics under multifold testing objectives.} Evaluating selection metrics under multiple objectives is essential for understanding their practical effectiveness, yet this aspect is missing in existing studies. We provide a unified evaluation benchmark with statistical analysis (Non-Parametric Scott-Knott Effect Size Difference test \cite{tantithamthavorn2016empirical}) that compares 15 metrics across 30 settings under four evaluation criteria (\#Mis., \#Clu., AE\%, Acc.\%). 

\textit{RQ3. Performance of test selection metrics under different selection budgets.}
This analyzes the stability and scalability of the studied metrics under varying labeling budgets, which is a critical factor when practitioners face strict constraints on labeling resources. To answer this, we provide heatmap plots demonstrating the ranking of each metric across different budgets for each evaluation criterion. 

\textit{RQ4. Time efficiency of test selection metrics.}
Efficient test selection is also an important aspect to consider when deploying the metrics in practice. In this question, we evaluate the time efficiency of all studied metrics. 

%% file: Experimental_setup.tex
\section{Experimental Setup}
In this section, we provide a summary of the test selection metrics investigated in this study, as well as the selection budgets. Then, we describe datasets and DNNs used in the experiments.

\begin{table}[!htbp]
\small
\setlength\extrarowheight{-1.0pt}
    \centering
    \caption{Summary of 15 test selection metrics studied in this paper. `Type' indicates the most suitable type the metric belongs to based on its characteristics. `C' and `R' denote whether this metric is applicable to classification and regression tasks, respectively. Under the `Obj.', we have listed which testing objective this metric has been studied in the literature: 1: fault detection, 2: performance estimation, 3: retraining guidance. `Year' denotes the time the metric was proposed.}
    \resizebox{0.8\textwidth}{!}{\begin{tabular}{c|c|c|c|c|c|c}
    \toprule
        Metrics  & Type & Description & C & R & Obj. & Year\\
    \midrule
     Gini \cite{feng2020deepgini} & Uncertainty & Select the most uncertain inputs & \cmark & \xmark & 1, 3 & 2020 \\
\midrule
Ent \cite{wang2014new} & Uncertainty & Select the most uncertain inputs & \cmark & \xmark & 1 & 2014 \\
\midrule
NC \cite{pei2017deepxplore} & Diversity & Select inputs with maximum NC & \cmark & \cmark & 1, 3 & 2017 \\
\midrule
KMNC \cite{ma2018deepgauge} & Diversity & Select inputs with maximum KMNC & \cmark & \cmark & 1 & 2018 \\
\midrule
GD \cite{aghababaeyan2023black} & Diversity & Select test suite with maximum diversity & \cmark & \cmark & 1 & 2023 \\
\midrule
STD \cite{aghababaeyan2023black} & Diversity & Select test suite with maximum diversity & \cmark & \cmark & 1 & 2023 \\
\midrule
LSA \cite{kim2019guiding} & Surprise & Select inputs with maximum surprise & \cmark & \cmark & 1, 3 & 2019 \\
\midrule
DSA \cite{kim2019guiding} & Surprise & Select inputs with maximum surprise & \cmark & \xmark & 1, 3 & 2019 \\
\midrule
CES \cite{li2019boosting} & Sampling & Select inputs that guarantee distribution similarity & \cmark & \cmark & 2 & 2019 \\
\midrule
PACE \cite{chen2020practical} & Clustering & Select representative inputs from each cluster & \cmark & \cmark & 2 & 2020 \\
\midrule
DR \cite{zhou2020cost} & Hybrid & Select inputs with multi-objective optimization & \cmark & \cmark & 2 & 2020 \\
\midrule
EST \cite{guerriero2021operation} & Sampling & Select inputs with adaptive sampling & \cmark & \cmark & 1, 2 & 2021 \\
\midrule
MCP \cite{shen2020multiple} & Clustering & Select inputs at decision boundary areas. & \cmark & \xmark & 3 & 2020 \\
\midrule
DAT \cite{hu2022empirical} & Hybrid & Selects uncertain ID inputs, randomly selects OOD inputs & \cmark & \xmark & 3 & 2022 \\
\midrule
Rand & Sampling & Randomly select inputs without replacement & \cmark & \cmark & NA & NA \\
     \midrule
    \end{tabular}}
    \label{tab:summary metrics}
\end{table}

\subsection{Test Selection Metrics} \label{sec: test selection metrics}
In this paper, we study 15 test selection metrics, as summarized in Table \ref{tab:summary metrics}. The year of proposed metrics ranges from 2014 to 2023, including classic metrics (e.g., NC in 2017 \cite{pei2017deepxplore}) and state-of-the-art metrics (e.g., GD and STD in 2023 \cite{aghababaeyan2023black}). Five metrics (Gini \cite{feng2020deepgini}, Ent \cite{wang2014new}, DSA \cite{kim2019guiding}, MCP \cite{shen2020multiple}, DAT \cite{hu2022empirical}) rely on the softmax confidence score that is not available in regression models, therefore, we only report their performance on classification tasks. As shown in the table, these 15 metrics have been studied for only one or two testing objectives, e.g., CES \cite{li2019boosting}, PACE \cite{chen2020practical}, and DR \cite{zhou2020cost} investigate their effectiveness for performance estimation only. However, how they perform under other important objectives is still underexplored. We also find that metrics studied for performance estimation lack analyzing their performance on OOD scenarios, i.e., most of them evaluate on the original ID testing set only. This paper provides an evaluation benchmark that analyzes 15 widely used metrics on three important testing objectives on five different OOD shifts, providing insightful findings and recommendations for researchers and practitioners in the SE community.  For implementation details of these metrics (e.g., layer selection for LSA, OOD detectors for DAT, auxiliary variables for EST), please refer to our online repository due to limited space.
Similar to existing works \cite{chen2020practical, hu2022empirical}, we select four different sizes of test suites to assess the effectiveness of the metrics, i.e., our studied \textbf{selection budgets are 50, 100, 150, and 200}.

\subsection{Datasets and DNNs}
As shown in Table \ref{tab:dataset_summary}, we design 30 experiment settings, which vary in the testing set, the DNN models, and the OOD types. The testing sets cover three different data types (image, text, and Android packages), with both classification (MNIST \cite{lecun1998mnist}, AndroZoo \cite{allix2016androzoo}, and IMDb \cite{maas2011learning}) and regression (Udacity \cite{udacity_dataset}) tasks. The DNN models are constructed with different structures, thus different accuracies on the testing sets (the column `Perf.'). Five different distribution shifts are used in the experiments: corrupted, adversarial, label, temporal, and natural shifts (the column `OOD Type'). Testing sets with distribution shifts are constructed with 50\% ID samples and 50\% OOD samples (see Section \ref{sec:OOD data construction} for details). As a result, we have a total of 10 $\times$ 4 $\times$ 30 + 5 $\times$ 4 $\times$ 22 combinations, summing up to 1,640 unique scenarios.

\begin{table*}[!htbp]
\setlength\extrarowheight{-0.5pt}
\centering
\small
\caption{Summary of all 30 experiment settings used in our evaluation. `\#Params' denotes the number of parameters in the model. `Perf.' indicates the model accuracy on the corresponding testing set. `\#Tests' is the number of samples in the testing set. }
\resizebox{0.8\textwidth}{!}{
\begin{tabular}{c|l|l|l|l|r|l}
\toprule
\textbf{ID} & \textbf{Testing Set} & \textbf{Model} & \textbf{\#Params} & \textbf{Perf.} & \textbf{\#Tests} & \textbf{OOD Type} \\

\midrule
1  & \multirow{3}{*}{MNIST \cite{lecun1998mnist}}      & LeNet-1 \cite{lecun1998gradient}             &    7,206   &    94.86\%   & 10,000  & Original \\
2  &       & LeNet-4      \cite{lecun1998gradient}       &    77,998   &  96.79\%     & 10,000  & Original \\
3  &       & LeNet-5  \cite{lecun1998gradient}          &   89,698    &   98.68\%    & 10,000  & Original \\
\midrule
4  & \multirow{4}{*}{Udacity \cite{udacity_dataset}}    & Dave2V1  \cite{bojarski2016end}           &   2,116,983    &    96.35\%\footnote{we use 1- MSE as the accuracy}   & 5,614   & Original \\
5  &     & Dave2V2  \cite{Dave2v2}   &  2,116,983      &    95.67\%         & 5,614   & Original \\
6  &     & Dave2V3   \cite{Dave2v3}          &    3,276,225   &  97.06\%     & 5,614   & Original \\
7  &     & Epoch    \cite{epoch_model}           &  18,969,665     &  98.41\%     & 5,614   & Original \\ 
\midrule
8  & \multirow{2}{*}{AndroZoo \cite{allix2016androzoo}}    & DeepDrebin   \cite{li2021can}       & 2,404,802 & 99.23\% & 21,336  & Original \\
9 & & BasicDNN \cite{li2020adversarial} & 1,626,081 & 99.22\% & 21,336 & Original \\
\midrule
10 & \multirow{4}{*}{IMDb \cite{maas2011learning}}       & Linear \cite{hendrycks2016baseline} &  640,033     &  87.48\%     & 25,000  & Original \\
11 &        & LSTM   \cite{LSTMGRUimdb}       &   692,785    &   85.47\%    & 25,000  & Original \\
12 &        & GRU    \cite{LSTMGRUimdb}       &   680,753    &   84.68\%    & 25,000  & Original \\
13 &        & Transformer \cite{Transformerimdb} &    653,566   &  87.57\%     & 25,000  & Original \\
\midrule
14 & MNIST-C  \cite{mu2019mnist}   & LeNet-5   \cite{lecun1998gradient}          &  89,698     & 92.07\%   
& 10,000  & Corrupted covariate shift \\
15 & MNIST-Adv & LeNet-5 \cite{lecun1998gradient}  &  89,698   &   50.06\%    & 10,000  & Adversarial covariate shift \\
16 & MNIST-label & LeNet-5 \cite{lecun1998gradient} &    89,698 &   98.97\%    & 10,000  & Label shift \\
17 & MNIST-EMNIST & LeNet-5  \cite{lecun1998gradient}    &   89,698  &   92.76\%    & 10,000  & Natural covariate shift \\
\midrule
18 & Udacity-C  & Epoch   \cite{epoch_model} &  18,969,665  &   97.19\%    & 5,614   & Corrupted covariate shift \\
19 & Udacity-Adv & Epoch \cite{epoch_model}  &  18,969,665     &   56.05\%    & 5,614   & Adversarial covariate shift \\
20 & Udacity-label & Epoch \cite{epoch_model}  &  18,969,665     &    95.85\%   & 5,614   & Label shift \\
21 & Udacity-Dave & Epoch \cite{epoch_model}   &    18,969,665   &   98.44\%    & 5,614   & Natural covariate shift \\
\midrule
22 & AndroZoo-2018 & DeepDrebin  \cite{li2021can}      & 2,404,802  & 96.33\% & 16,000 
& Temporal covariate shift \\
23 & AndroZoo-2019 & DeepDrebin  \cite{li2021can}      & 2,404,802  & 96.36\% & 16,000  & Temporal covariate shift \\
24 & AndroZoo-Adv & DeepDrebin  \cite{li2021can}      & 2,404,802 & 57.28\% & 21,336  & Adversarial covariate shift \\
25 & AndroZoo-label & DeepDrebin  \cite{li2021can}     & 2,404,802 & 97.84\% & 21,336  & Label shift \\
26 & AndroZoo-Drebin & DeepDrebin  \cite{li2021can}    & 2,404,802 & 84.48\% & 21,336  & Natural covariate shift \\ 
\midrule
27 & IMDb-C  \cite{weiss2022simple}   & Transformer  \cite{Transformerimdb}       &  653,566 & 74.99\% & 25,000  & Corrupted covariate shift \\
28 & IMDb-Adv  & Transformer   \cite{Transformerimdb}      &  653,566     &  42.89\%     & 25,000  & Adversarial covariate shift \\
29 & IMDb-label & Transformer \cite{Transformerimdb}        & 653,566 & 88.02\% & 25,000  & Label shift \\
30 & IMDb-Customer & Transformer\cite{Transformerimdb} &  653,566   & 78.90\% & 7,550  & Natural covariate shift \\
\bottomrule
\end{tabular}}
\label{tab:dataset_summary}
\end{table*}

%% file: Results.tex
\section{Results}
This section presents the results of experiments we used to answer the research questions.

\input{RQ1}
\input{RQ2}

\input{RQ3}
\input{RQ4}

%% file: RQ1.tex
\subsection{RQ1.1. Selection of optimal clustering pipelines.} \label{sec:RQ1.1.cluster}

\subsubsection{The clustering quality.}
 We conduct preliminary clustering experiments with different candidate techniques for each step on the mispredicted samples from the original testing set of each dataset, averaging over all studied models. Recall that the clustering pipeline contains three steps: feature extraction (FE), dimensionality reduction (DR), and clustering (CA to denote clustering algorithms). We use Silhouette \cite{rousseeuw1987silhouettes} and DBCV \cite{moulavi2014density} scores for evaluation, where both range from -1 to 1. A higher value indicates a better clustering result. For the AndroZoo dataset, we assess DeepDrebin and BasicDNN \cite{li2021can, li2020adversarial} for FE, where we use the first hidden layer for both \cite{yosinski2014transferable}, three techniques for DR: Principal Component Analysis (PCA) \cite{pearson1901liii}, Uniform Manifold Approximation and Projection (UMAP) \cite{mcinnes2018umap}, and Gaussian Random Projection (GRP) \cite{bingham2001random}, and four CA: K-Means \cite{macqueen1967some}, Hierarchical Agglomerative Clustering (HAC) \cite{king2013cluster}, Density-based spatial clustering of applications with noise (DBSCAN) \cite{ester1996density}, and Hierarchical DBSCAN (HDBSCAN) \cite{mcinnes2017hdbscan}. We perform a similar procedure for the IMDb dataset, which tests pre-trained BERT \cite{devlin2019bert}, RoBERTa \cite{liu2019roberta}, and ELECTRA \cite{clark2020electra} for FE (we use the penultimate layer for all \cite{devlin2019bert}), the same candidates for DR and CA as in AndroZoo. For a fair pipeline comparison, we use our optimally tuned hyperparameter values (see online repository for details). To streamline our analysis, we conduct ablation studies by varying techniques in one step while holding the other two constant. For instance, when evaluating FE candidates, we fix UMAP for DR and DBSCAN for CA for all datasets, following the approach recommended by \cite{attaoui2024supporting}. Then, we fix the identified optimal FE method and DBSCAN for CA when ablating DR candidates. Finally, we fix the optimal FE and DR methods when ablating CA candidates. Based on the Silhouette and DBCV scores in Table \ref{tab: clustering-scores}, the optimal pipelines are \textbf{(DeepDrebin, UMAP, DBSCAN)} for the AndroZoo dataset and \textbf{(RoBERTa, UMAP, DBSCAN)} for IMDb. Note that for image datasets (MNIST and Udacity), we directly use the best pipeline \textbf{(ResNet-50 \cite{he2016deep}, UMAP, DBSCAN)} validated by \cite{attaoui2024supporting}, which achieves average scores of 0.69 and 0.47 for MNIST and Udacity, respectively.
\begin{table}[!htbp]
\caption{The Silhouette and DBCV results of candidate pipelines for AndroZoo and IMDb datasets, average over all studied models. `Deep': DeepDrebin; `Basic': BasicDNN; `RBT': RoBERTa; `BT': BERT; `ET': ELECTRA; `DB': DBSCAN; `HDB': HDBSCAN, `KM': K-Means.}
    \centering
    \begin{subtable}{0.48\textwidth}  
    \centering
    \caption{AndroZoo. }
    \scriptsize
\addtolength{\tabcolsep}{-3pt}
\setlength\extrarowheight{-0.5pt}
    \begin{tabular}{c c c |ccc|cccc}
    \toprule
     &  \multicolumn{2}{c|}{\textbf{FE}} & \multicolumn{3}{c|}{\textbf{DR}} & \multicolumn{4}{c}{\textbf{CA}} \\
      \textbf{Score} & Deep & Basic & UMAP & PCA & GRP & DB & HDB & HAC & KM \\
    \textbf{Sil.} & \cellcolor{gray!25}0.83 & 0.58 & \cellcolor{gray!25}0.83 & 0.06 & 0.04 & \cellcolor{gray!25}0.83& 0.56 & 0.33 & 0.51\\
    \textbf{DBCV} & \cellcolor{gray!25}0.91 & 0.75 & \cellcolor{gray!25}0.91 & -0.47 & -0.46 & \cellcolor{gray!25}0.91 & 0.32 & 0.24 & 0.02\\
    \textbf{Avg} & \cellcolor{gray!25}0.87 & 0.67 & \cellcolor{gray!25}0.87 & -0.21 & -0.21 & \cellcolor{gray!25}0.87 & 0.44 & 0.29 & 0.27  \\
    \bottomrule
    \end{tabular}
    \label{tab:clustering-pipeline-androzoo}
    \end{subtable}
     \begin{subtable}{0.453\textwidth}
    \centering
    \caption{IMDb.}
    \scriptsize
\addtolength{\tabcolsep}{-3pt}
\setlength\extrarowheight{-0.5pt}
    \begin{tabular}{c c c |ccc|cccc}
    \toprule
      \multicolumn{3}{c|}{\textbf{FE}} & \multicolumn{3}{c|}{\textbf{DR}} & \multicolumn{4}{c}{\textbf{CA}} \\
   RBT & BT & ET &  UMAP & PCA & GRP & DB & HDB & HAC & KM  \\
       \cellcolor{gray!25}0.91 & 0.16 & 0.43 & \cellcolor{gray!25}0.91 & 0.25 & 0.26 & \cellcolor{gray!25}0.91 & 0.87 & 0.32 & 0.17 \\
        \cellcolor{gray!25}0.93 & 0.06 & 0.43 & \cellcolor{gray!25}0.93 & 0.01 & 0.01 & \cellcolor{gray!25}0.93 & 0.90 & 0.43 & -0.28 \\
     \cellcolor{gray!25}0.92 & 0.11 & 0.43 & \cellcolor{gray!25}0.92 & 0.13 & 0.14 & \cellcolor{gray!25}0.92 & 0.89 & 0.38 & -0.06 \\
        \bottomrule
    \end{tabular}
    \end{subtable}
\label{tab: clustering-scores}
\end{table}

\subsubsection{The cluster-to-fault correspondence.}
\begin{figure}[!htbp]
    \centering
    \begin{subfigure}{0.246\textwidth}
        \includegraphics[width=\textwidth]{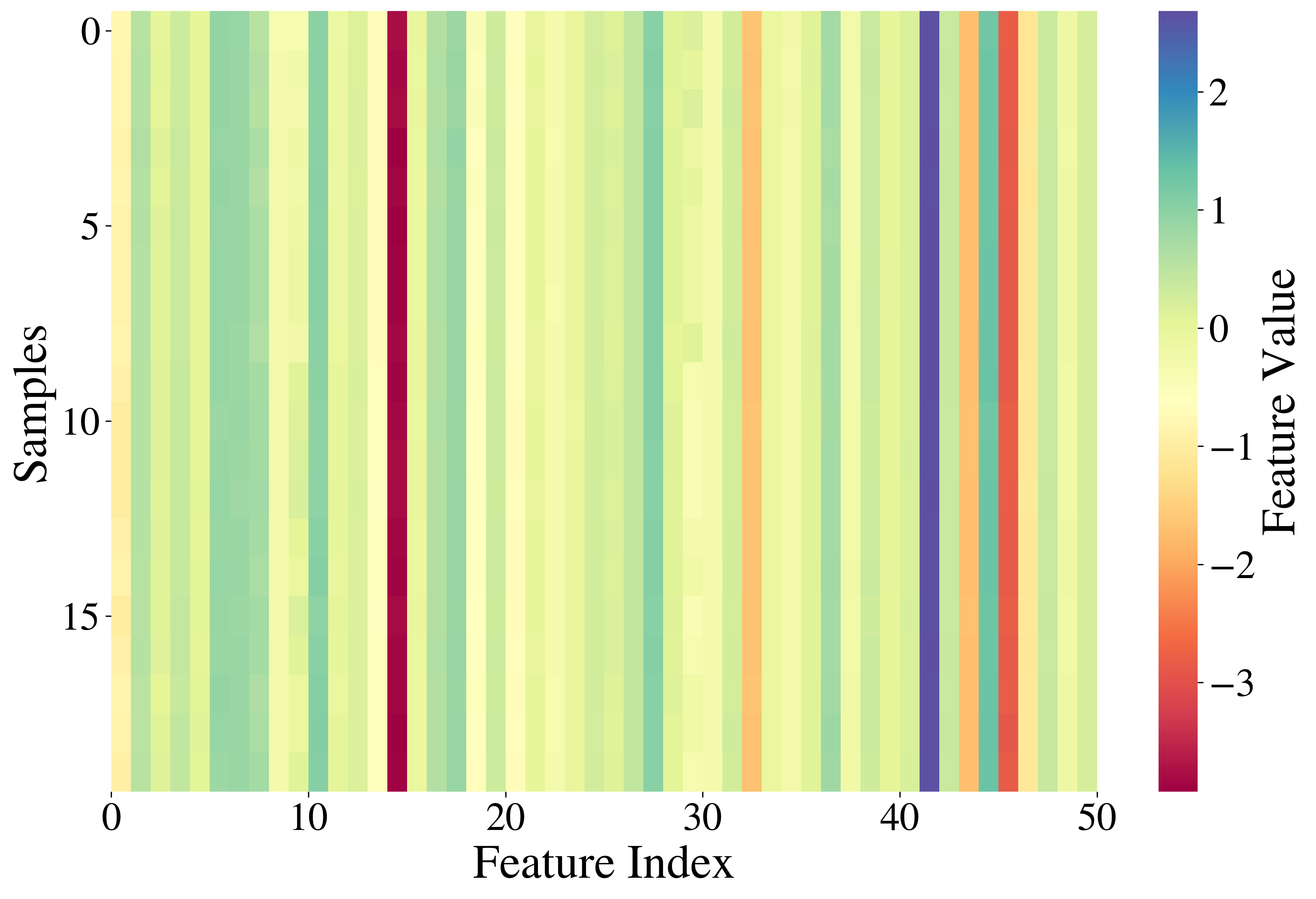}
        \caption{Udacity (best)}
        \label{fig:1a}
    \end{subfigure}
    \hfill
    \begin{subfigure}{0.246\textwidth}
        \includegraphics[width=\textwidth]{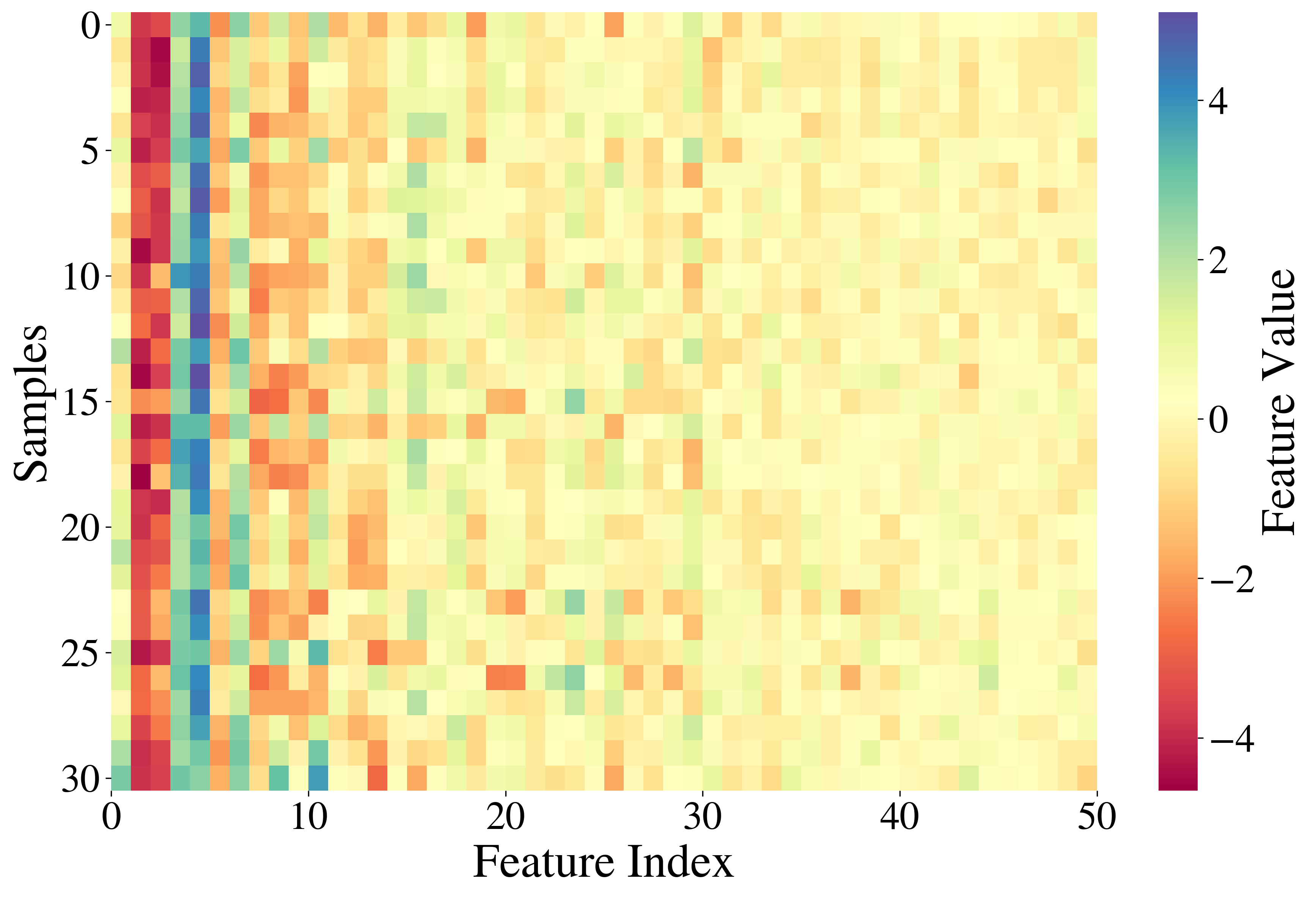}
        \caption{Udacity (other)}
        \label{fig:1b}
    \end{subfigure}
    \hfill
    \begin{subfigure}{0.246\textwidth}
        \includegraphics[width=\textwidth]{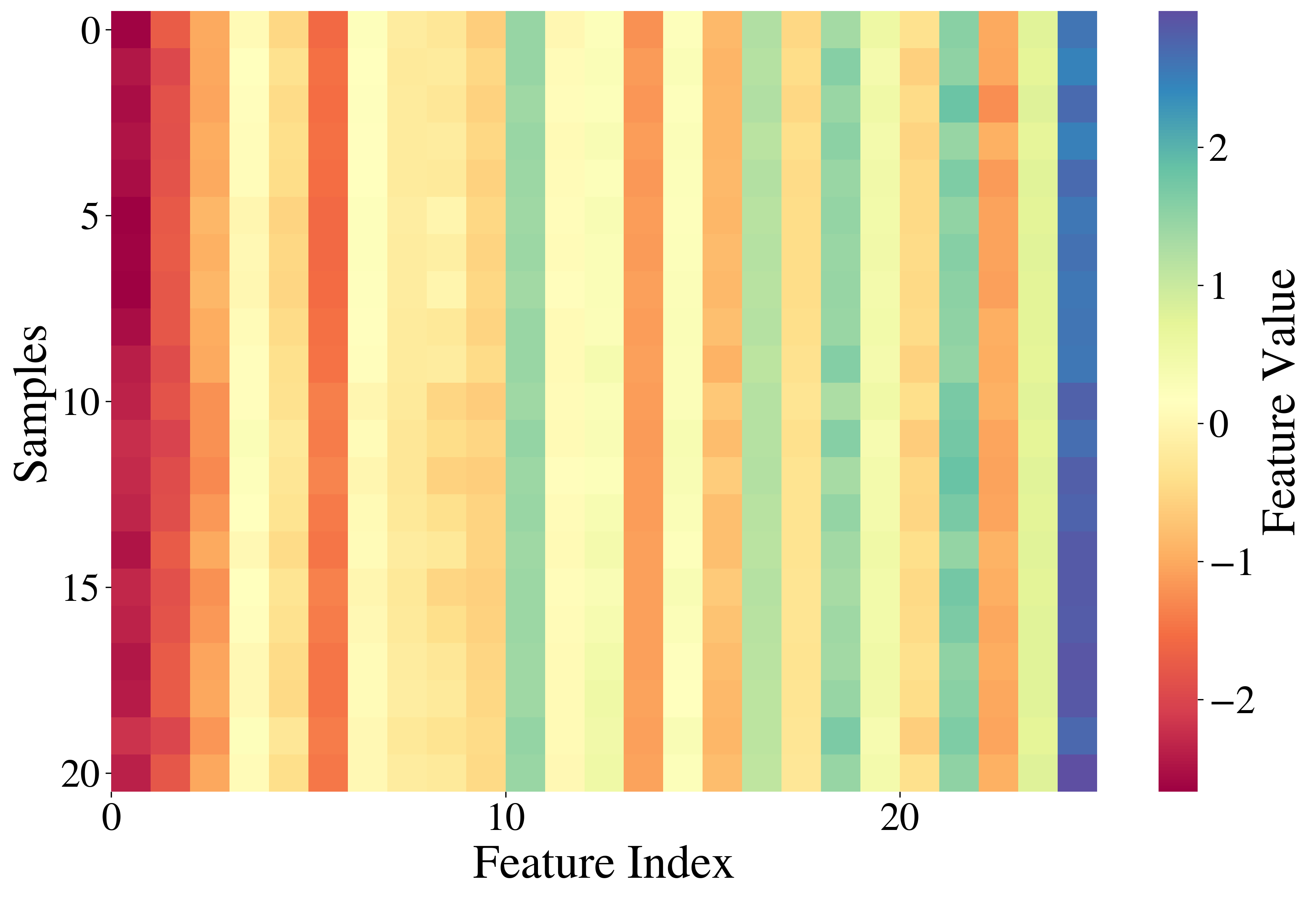}
        \caption{AndroZoo (best)}
        \label{fig:1c}
    \end{subfigure}
    \hfill
    \begin{subfigure}{0.245\textwidth}
        \includegraphics[width=\textwidth]{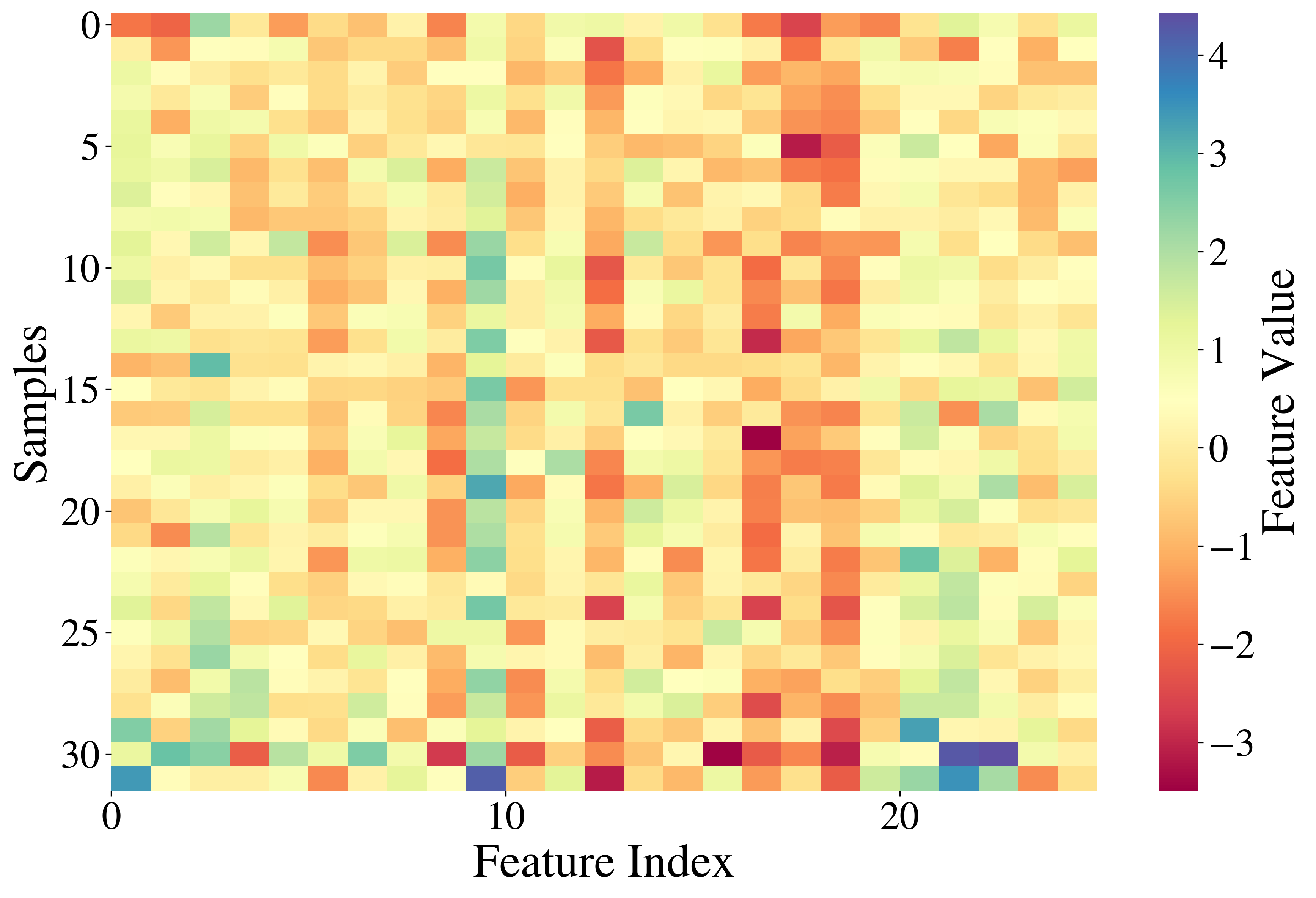}
        \caption{AndroZoo (other)}
        \label{fig:1d}
    \end{subfigure}
    \caption{Feature visualization of samples in a high-quality cluster (`best', Udacity: ResNet-50, UMAP, DBSCAN; Andro: DeepDrebin, UMAP, DBSCAN) and a low-quality cluster (`other', Udacity: ResNet-50, PCA, HAC; Andro: BasicDNN, GRP, K-Means).}
    \label{fig:heatmap-feature-inspection}
\end{figure}

To validate whether the resulting cluster from the best pipeline indeed represents a DNN fault, we conduct feature pattern inspection and cluster-specific retraining validation \cite{aghababaeyan2023black}. Figure \ref{fig:heatmap-feature-inspection} displays the heatmaps of a randomly selected cluster from the best clustering pipeline and a randomly selected cluster obtained from another pipeline. Due to limited space and similar patterns, we only show results of the Udacity and AndroZoo datasets (see online repository for MNIST and IMDb). Each row represents a single mispredicted sample, and each column represents a (reduced) feature. We colored the feature values using the spectral colormap. By comparing the `best' and `other' heatmaps, we can see that the high-quality cluster displays a uniform feature pattern across all mispredicted samples within this cluster, which suggests a shared underlying root cause of the fault.  On the other hand, the low-quality cluster demonstrates dispersed and inconsistent feature patterns across mispredicted samples within that cluster.

Table \ref{tab:quantitative-analysis-cluster-} demonstrates the results of cluster-specific retraining validation. Retraining is conducted using the 85\% samples from $C_i$ (we randomly pick three clusters from the best pipeline on one of the studied models). We validated and reported the accuracy improvement on the remaining 15\% samples of $C_i$ and on all samples in $C_{j \neq i}$. The results show that the retrained model is significantly more accurate on the cluster $C_i$ for which it was retrained, and less accurate on other clusters ($C_j, j \neq i$) for all datasets. Specifically, the differences in accuracy improvement are 46\%, 68\%, 49\%, and 63\% for MNIST, Udacity, AndroZoo, and IMDb, respectively. This indicates that retraining fixes the fault represented by $C_i$ but not the faults represented by $C_j$. Therefore, we can conclude that each cluster represents a unique fault and different clusters correspond to distinct faults. 
\begin{table}[!htbp]
    \centering
    \setlength\extrarowheight{-2.5pt}
    \footnotesize
    \caption{Cluster-specific retraining validation on all studied datasets.}
    \begin{tabular}{c|c c | cc | cc | cc}
    \toprule
    & \multicolumn{2}{c|}{MNIST} & \multicolumn{2}{c|}{Udacity} & \multicolumn{2}{c|}{AndroZoo} & \multicolumn{2}{c}{IMDb} \\
    Cluster ID $C_i$ & $C_i$ & $C_{j\neq i}$  & $C_i$ & $C_{j\neq i}$ & $C_i$ & $C_{j\neq i}$ & $C_i$ & $C_{j\neq i}$  \\
    \midrule
       Cluster 0 & \cellcolor{gray!25}72\% & 21\% & \cellcolor{gray!25}75\% & 19\% & \cellcolor{gray!25}71\% & 26\% & \cellcolor{gray!25}83\% & 12\%\\
       Cluster 1 & \cellcolor{gray!25}67\%  & 22\% & \cellcolor{gray!25}99\% & 18\% & \cellcolor{gray!25}86\% & 28\% & \cellcolor{gray!25}56\% & 23\% \\
       Cluster 2 & \cellcolor{gray!25}61\% & 20\% & \cellcolor{gray!25}88\% & 20\% & \cellcolor{gray!25}85\% & 42\% & \cellcolor{gray!25}98\% & 13\%  \\
       \midrule
       \textbf{Average} & \cellcolor{gray!25}67\% & 21\% & \cellcolor{gray!25}87\% & 19\% & \cellcolor{gray!25}81\% & 32\% & \cellcolor{gray!25}79\% & 16\%  \\
       \bottomrule
    \end{tabular}
    \label{tab:quantitative-analysis-cluster-}
\end{table}
\begin{tcolorbox}[colframe=red!40!black, sharp corners, leftrule={3pt}, rightrule={0pt}, toprule={0pt}, bottomrule={0pt}, left={2pt}, right={2pt}, top={3pt}, bottom={3pt}]
\textbf{Finding 1:} Heatmap visualization demonstrates the coherence of intra-cluster feature patterns from the best pipeline, and cluster-specific retraining further validates cluster-to-fault correspondence.
\end{tcolorbox}

\subsection{RQ1.2. Selection of optimal retraining processes.}
We report the results of type I and type II retraining for the Udacity and AndroZoo datasets, respectively. We adopt the recommended process (type II retraining) by Hu \textit{et~al.} \cite{hu2022empirical} for MNIST and IMDb datasets. Specifically, We conduct experiments on all budgets (50, 100, 150, 200), and report the average results over all applicable test selection metrics. The hyperparameters of retraining (e.g., learning rate, optimizer) are set in the same way as in the original training process. Results are illustrated in Table \ref{tab: retraining comparison}. For Udacity, in 20 out of 32 cases, type II retraining demonstrates a higher accuracy improvement. Specifically, type II retraining gives an improvement of 3.72\% on average, which is 2.06\% higher than type I retraining. We found that type II retraining gives outstanding results compared to type I when there is an adversarial shift. 
For AndroZoo, we observe an opposite trend, where type I outperforms type II retraining in 25 out of 28 cases by a margin of 0.95\% on average. We observe that the recommended retraining processes differ across data types. We analyze these differences from the perspectives of input geometry and feature characteristics. Udacity data are represented by spatially structured pixels with an input shape of (3,100,100), whereas AndroZoo data are represented with 10,000-dimensional binary vectors that are extremely sparse (around 99\% zeros) without semantic continuity, where a small pattern change may flip classification. In this case, Type I allows the model to zoom in on discriminative features of new data more than Type II. For Udacity, dense image features are consistent in semantics. Type II preserves a full manifold of visual contexts for stable model updates, however, under some certain cases (ID 7, 18, 20 in Table \ref{tab: retraining comparison Udacity}), Type I performs better since it allows the model to focus on the newly identified visual pattern.

\begin{table}[!htbp]
  \centering
  \caption{Comparison of type I and type II retraining processes for Udacity and AndroZoo datasets.}
  \begin{subtable}{0.45\textwidth} 
    \centering
    \setlength\extrarowheight{-1.5pt}
    \caption{Udacity.}
    \label{tab: retraining comparison Udacity}
    \resizebox{\textwidth}{!}{
    \begin{tabular}{c| cc cc cc cc} 
    \toprule
   &  \multicolumn{2}{c}{bg=50} & \multicolumn{2}{c}{bg=100} & \multicolumn{2}{c}{bg=150} & \multicolumn{2}{c}{bg=200} \\
    ID & Type I & II & I &  II & I &  II & I &  II \\
    \midrule
    4 & 1.6 & \cellcolor{gray!25}2.03 & 1.63 & \cellcolor{gray!25}2.07 & 1.68 & \cellcolor{gray!25}2.11 & 1.72 & \cellcolor{gray!25}2.16 \\
    5 & 0 & \cellcolor{gray!25}0.04 & 0 & \cellcolor{gray!25}0.04 & 0 & \cellcolor{gray!25}0.05 & 0 & \cellcolor{gray!25}0.05 \\
    6 & 0.77 & \cellcolor{gray!25}0.99 & 0.79 & \cellcolor{gray!25}1.02 & 0.79 & \cellcolor{gray!25}0.98 & 0.81 & \cellcolor{gray!25}0.92 \\
    7 & \cellcolor{gray!25}0.12 & -1.43 & \cellcolor{gray!25}0.12 & -1.48 & \cellcolor{gray!25}0.12 & -1.51 & \cellcolor{gray!25}0.12 & -1.52 \\
    18 & \cellcolor{gray!25}0.57 & 0.38 & \cellcolor{gray!25}0.57 & 0.34 & \cellcolor{gray!25}0.6 & 0.37 & \cellcolor{gray!25}0.63 & 0.39 \\
    19 & 7.43 & \cellcolor{gray!25}26.77 & 7.64 & \cellcolor{gray!25}27.46 & 8.02 & \cellcolor{gray!25}28.31 & 8.46 & \cellcolor{gray!25}29.24 \\
    20 & \cellcolor{gray!25}1.23 & -1.27 & \cellcolor{gray!25}1.26 & -1.28 & \cellcolor{gray!25}1.3 & -1.21 & \cellcolor{gray!25}1.35 & -1.09 \\
    21 & 0.98 & \cellcolor{gray!25}1.08 & 1 & \cellcolor{gray!25}1.1 & 1.03 & \cellcolor{gray!25}1.08 & 1.06 & \cellcolor{gray!25}1.07 \\
    \midrule
    \textbf{Avg.} &  1.59 & \cellcolor{gray!25}3.57 & 1.63 & \cellcolor{gray!25}3.66 & 1.69 & \cellcolor{gray!25}3.77 & 1.77 & \cellcolor{gray!25}3.9 \\
    \bottomrule
    \end{tabular}}
    \end{subtable}
    \hspace{3mm}
  \begin{subtable}{0.45\textwidth} \label{tab: retraining comparison AndroZoo}
    \centering
    \setlength\extrarowheight{-1.5pt}
    \caption{AndroZoo.}
    \resizebox{\textwidth}{!}{
    \begin{tabular}{c | cccccccc}
    \toprule
    &  \multicolumn{2}{c}{bg=50} & \multicolumn{2}{c}{bg=100} & \multicolumn{2}{c}{bg=150} & \multicolumn{2}{c}{bg=200} \\
    ID & Type I & II & I &  II & I &  II & I &  II \\
    \midrule
    8  & \cellcolor{gray!25}-0.13 & -0.24 & \cellcolor{gray!25}-0.13 & -0.35 & \cellcolor{gray!25}-0.23 & -0.89 & \cellcolor{gray!25}-0.23 & -0.5 \\
    9  & \cellcolor{gray!25}0.01 & -0.2 & \cellcolor{gray!25}0.02 & -0.2 & \cellcolor{gray!25}-0.02 & -0.34 & \cellcolor{gray!25}-0.04 & -0.51 \\
    22 & \cellcolor{gray!25}-2.38 & -3.18 & \cellcolor{gray!25}-1.12 & -2.42 & \cellcolor{gray!25}-1.76 & -4.46 & \cellcolor{gray!25}-3.87 & -5.14 \\
    23 & -2.22 & \cellcolor{gray!25}-2.12 & \cellcolor{gray!25}-0.95 & -2.41 & \cellcolor{gray!25}-1.73 & -4.65 & \cellcolor{gray!25}-4.01 & -4.7 \\
    24 & \cellcolor{gray!25}4.14 & 1.17 & \cellcolor{gray!25}4.82 & 3.11 & \cellcolor{gray!25}4.48 & 1.74 & \cellcolor{gray!25}3.75 & 0.26 \\
    25 & \cellcolor{gray!25}0.04 & -0.36 & \cellcolor{gray!25}0.19 & -0.34 & \cellcolor{gray!25}0.06 & -0.56 & \cellcolor{gray!25}-0.25 & -0.49 \\
    26 & 7.78 & \cellcolor{gray!25}8.72 & 8.06 & \cellcolor{gray!25}8.6 & \cellcolor{gray!25}8.18 & 7.68 & \cellcolor{gray!25}8 & 6.75 \\
    \hline
    \textbf{Avg.} &  \cellcolor{gray!25}1.03 & 0.54 & \cellcolor{gray!25}1.56 & 0.86 & \cellcolor{gray!25}1.28 & -0.21 & \cellcolor{gray!25}0.48 & -0.62 \\
    \hline
  \end{tabular}}
  \end{subtable}
\label{tab: retraining comparison}
\end{table}

\begin{tcolorbox}[colframe=red!40!black, sharp corners, leftrule={3pt}, rightrule={0pt}, toprule={0pt}, bottomrule={0pt}, left={2pt}, right={2pt}, top={3pt}, bottom={3pt}]
\textbf{Finding 2:} For Udacity, type II outperforms type I retraining by approximately 3.06\% on average. We observe an opposite trend for AndroZoo data, where type I outperforms type II by 0.95\% on average. 
\end{tcolorbox}

%% file: RQ2.tex
\input{RQ2_tables.tex}
\subsection{RQ2. Performance of test selection metrics under multifold testing objectives.}
\subsubsection{Statistical significance} We conduct a statistical analysis to provide an overall ranking of all 15 metrics examined in this paper. To determine the statistical significance of the observed performance differences among multiple test selection metrics, we employ the Non-Parametric Scott-Knott Effect Size Difference (NPSK) test \cite{tantithamthavorn2016empirical}, which is a multiple comparison approach that leverages a hierarchical clustering to partition the set of median values of techniques into statistically distinct groups with a non-negligible difference. The NPSK does not require the assumptions of normal distributions, homogeneous distributions, and the minimum sample size. Different groups exhibit statistically significant differences at the predetermined significance level of 0.05 ($\alpha$ = 0.05). The NPSK ensures the magnitude of the difference between metrics within each group is not statistically significant, and the magnitude of the difference between metrics located in different groups is statistically significant. We use the same color for metrics within the same group and different colors for different groups. We utilize three distinct colors to underscore the statistical differences among 15 test selection metrics for each setting in terms of four evaluation criteria. Darker colors represent a superior performance. Specifically, the darkest blue denotes the first group, moderate blue indicates the second group, and the lightest blue represents the third group.  If there are at least five groups, the results of the fourth group will be bolded while maintaining a white cell background. We use results collected from 4 different budgets for statistical test.

\subsubsection{Key results} Tables \ref{tab:main-results1}, \ref{tab:main-results2} show the results of 15 studied test selection metrics on 4 evaluation criteria across four benchmark datasets with five types of OOD scenarios. The average results are computed over all applicable settings. We perform within-task-type comparison for each objective.

\textbf{Fault Detection.} The effectiveness of a test suited to satisfy this testing objective is demonstrated by two criteria: number of mispredictions (\#Mis) and number of clusters (\#Clu.). For classification tasks (22 settings in total), the test suite selected by DSA detects the highest number of misclassifications and is located in the best statistical group in 21 out of 22 settings, with an average of 105.11 misclassifications detected. Gini and Ent appear in the best statistical group in the left case, detecting an average of 59.14 and 58.60 misclassifications. Meanwhile, DSA detects the highest number of clusters and is located in the best statistical group in 12 out of 22 settings (10.16 clusters on average), with DAT, Gini, Ent, and DR locating in the best statistical group in 7, 3, 3, and 1 settings, respectively. For the regression task (Udacity dataset, 8 settings in total), recall that we report the average number of mispredictions and clusters over 11 thresholds for criteria `\#Mis.' and `\#Clu.', respectively (Section \ref{sec:study-design-fault-detection}). Test suites selected by NC are located in the best statistical group in `\#Mis.' and `\#Clu.' in 6 settings for each, detecting 66.09 mispredictions and 11.92 clusters on average.

\begin{tcolorbox}[colframe=red!40!black, sharp corners, leftrule={3pt}, rightrule={0pt}, toprule={0pt}, bottomrule={0pt}, left={2pt}, right={2pt}, top={3pt}, bottom={3pt}]
\textbf{Finding 3:} With fault detection as the testing objective, in general, we recommend using DSA for classification tasks and NC for regression tasks. 
\end{tcolorbox}

\textbf{Performance Estimation.} This objective is assessed by AE\%. For classification tasks, Rand achieves the most accurate estimation with statistical significance in 6 out of 22 settings, followed by DAT located in the best statistical group in 4 settings. On average, Rand demonstrates the lowest error (1.97\%), followed by STD and GD, which give AE\% of 2.16\% and 2.19\%, averaging over all 22 classification settings. We note that metrics that cover different aspects of test suites can ensure the distribution similarity between the selected set and the whole testing set, which gives a more accurate estimation. However, in our studied settings, metrics specifically designed for performance estimation (CES, PACE, EST, and DR) cannot even outperform Rand, where EST demonstrates a large error of 22.68\%. It is also worth noting that DSA gives the poorest performance, with the largest average error (73.07\%). Similarly, uncertainty-based metrics, Gini and Ent also demonstrate poor performance. For regression tasks, LSA, Rand, and EST appear in the best statistical group in 3, 2, and 2 out of 8 settings, leading to an error value of 6.86\%, 1.6\%, and 1.95\%, respectively. NC gives the worst performance, with an average AE\% of 20.01\%, followed by KMNC and CES, with error values of 13.36\% and 10.03\%, respectively.

 \begin{tcolorbox}[colframe=red!40!black, sharp corners, leftrule={3pt}, rightrule={0pt}, toprule={0pt}, bottomrule={0pt}, left={2pt}, right={2pt}, top={3pt}, bottom={3pt}]
\textbf{Finding 4:} Metrics (Rand, GD, STD) that encourage test diversity outperform metrics (CES, PACE, EST, and DR) that are specifically designed for performance estimation. EST (22.68\%) and CES (10.03\%) give a very inaccurate estimation in classification and regression tasks, respectively.
\end{tcolorbox}

\textbf{Retraining Guidance.}
Overall, retraining with test suites selected from datasets containing original inputs only (e.g., MNIST, MNIST-label) or mild distribution shifts (e.g., Udacity-C, IMDb-Customer) produces a small (e.g., ID 3, 18, 25)  or even negative (e.g., ID 7, 8, 9, 11, 13, 20, 30) performance improvement. In these settings, all metrics achieve similar results because the model is already well-trained with the input distribution, and retraining on the selected set does not provide new information. In contrast, datasets containing adversarial OOD shifts (e.g., Udacity-Adv, MNIST-Adv) show significantly larger retraining gains (e.g., ID 15, 19, 28). For classification tasks, KMNC (4.37\%) gives the highest average improvement, followed by GD (4.31\%) and PACE (4.15\%). They are located in the best statistical group in 2, 1, and 8 settings, respectively. DSA (-0.28\%) gives the poorest performance. Metrics (MCP, DAT) that are designed solely for this objective achieves an average improvement of 2.85\% and 4.01\%, respectively. For regression tasks, PACE (5.25\%) gives the largest average improvement, followed by STD (5.24\%).

\begin{tcolorbox}[colframe=red!40!black, sharp corners, leftrule={3pt}, rightrule={0pt}, toprule={0pt}, bottomrule={0pt}, left={2pt}, right={2pt}, top={3pt}, bottom={3pt}]
\textbf{Finding 5:} KMNC and PACE demonstrate the highest retraining improvement for classification and regression, respectively, outperforming metrics (MCP, DAT) specifically designed for this objective. 
\end{tcolorbox}

\textbf{OOD sensitivity analysis.} 
We analyze how the effectiveness of test selection metrics varies across different types of OOD shifts (corrupted, adversarial, label, temporal, and natural shifts). 
Our results demonstrate that, with a few exceptions, the relative performance rankings of test selection metrics remain largely stable across different OOD scenarios. For example, DSA and NC stably reaches top performance in fault detection for classification and regression tasks across different OOD types, respectively. However, we note that CES (1.06\%) gives the best performance estimation in label shifts for classification tasks (3 settings), followed by Rand (1.59\%). For the regression task (1 setting), EST performs the best in adversarial (1.08\%) and corrupted (1.27\%) shifts, followed by STD (2.5\%) and Rand (2.04\%), respectively. In general, metrics (Rand, GD, STD) that encourage test diversity still achieve outstanding performance in performance estimation. Overall, the effectiveness of test selection metrics is not highly sensitive to the type of OOD shifts, and the choice of metric can be made without needing to re-adjust for different OOD scenarios.

\begin{tcolorbox}[colframe=red!40!black, sharp corners, leftrule={3pt}, rightrule={0pt}, toprule={0pt}, bottomrule={0pt}, left={2pt}, right={2pt}, top={3pt}, bottom={3pt}]
\textbf{Finding 6:} Overall, there is no clear fluctuations in metrics' performance under different OOD scenarios.  
\end{tcolorbox}

%% file: RQ2_tables.tex
\begin{table*}[!htbp]
\centering
\scriptsize
\addtolength{\tabcolsep}{-3.5pt}
\setlength\extrarowheight{-0.5pt}
\caption{Performance of 15 test selection metrics across 30 settings under four evaluation criteria: \#Mis. ($\uparrow$), \#Clu. ($\uparrow$), AE\% ($\downarrow$), and Acc.\% ($\uparrow$). Results are averaged over 4 budgets. Darker cells signify superior performance, while various colors denote the statistical significance among the 15 studied test selection metrics for each ID and criterion (Non-Parametric Scott-Knott ESD test with
a \textit{p}-value $\leq 0.05$) (ID: 1-15).}
\resizebox{0.9\textwidth}{!}{
\begin{tabular}{c|c|cccccccccccccccc}
\toprule
& \textbf{ID} & \textbf{Eval.} & \textbf{Rand} & \textbf{Gini} & \textbf{Ent} & \textbf{NC} & \textbf{KMNC} & \textbf{GD} & \textbf{STD} & \textbf{LSA} & \textbf{DSA} & \textbf{CES} & \textbf{PACE}  & \textbf{EST} & \textbf{DR} & \textbf{MCP} & \textbf{DAT} \\
\midrule
\multirow{12}{*}{\textbf{MNIST}} & \multirow{4}{*}{1} & \#Mis.   &                                  6.75 &  \cellcolor{MidnightBlue!40!white}68.75 & \cellcolor{MidnightBlue!40!white}64.25 &                                     5.0 &                                    3.75 &                                    7.0 &                                   4.75 &                                   6.25 &  \cellcolor{MidnightBlue!10!white}45.25 &                                   6.75 &                                   5.25 &                          \textbf{38.0} &                                    8.0 &  \cellcolor{MidnightBlue!20!white}57.25 &                                   3.25 \\

 &  & \#Clu.   &                                   1.5 &  \cellcolor{MidnightBlue!40!white}12.75 & \cellcolor{MidnightBlue!40!white}12.25 &                                    1.25 &                                     1.0 &                          \textbf{1.75} &                                    0.5 &                                    1.0 &   \cellcolor{MidnightBlue!20!white}7.50 &                          \textbf{1.75} &                                    1.0 &   \cellcolor{MidnightBlue!10!white}7.0 &                                    1.5 &  \cellcolor{MidnightBlue!40!white}12.00 &                                   0.50 \\
\end{tabular}}
\label{tab:main-results1}
\end{table*}

\begin{table*}[!htbp]
\centering
\scriptsize
\addtolength{\tabcolsep}{-3.5pt}
\setlength\extrarowheight{-0.5pt}
\caption{Performance of 15 test selection metrics across 30 settings under four evaluation criteria: \#Mis. ($\uparrow$), \#Clu. ($\uparrow$), AE ($\downarrow$), and Acc.\% ($\uparrow$) results are averaged over 4 budgets (ID: 16-30).}
\resizebox{0.9\textwidth}{!}{
\begin{tabular}{c|c|cccccccccccccccc}
\toprule
  &  \textbf{ID} & \textbf{Eval.} & \textbf{Rand} & \textbf{Gini} & \textbf{Ent} & \textbf{NC} & \textbf{KMNC} & \textbf{GD} & \textbf{STD} & \textbf{LSA} & \textbf{DSA} & \textbf{CES} & \textbf{PACE} & \textbf{EST} & \textbf{DR} & \textbf{MCP} & \textbf{DAT} \\
\midrule
\multirow{4}{*}{\textbf{MNIST-label}} & \multirow{4}{*}{16} & \#Mis.   &                                  1.75 &  \cellcolor{MidnightBlue!20!white}46.75 &  \cellcolor{MidnightBlue!20!white}47.0 &                                     1.0 &                                     1.5 &                                    1.5 &                                   1.75 &                                    1.0 &  \cellcolor{MidnightBlue!40!white}55.25 &                                    0.5 &                                    3.5 &                         \textbf{33.75} &                                    1.0 &  \cellcolor{MidnightBlue!10!white}42.25 &                                   1.75 \\

 &  & \#Clu.   &                         \textbf{0.25} &   \cellcolor{MidnightBlue!40!white}10.5 &  \cellcolor{MidnightBlue!40!white}11.0 &                                     0.0 &                                     0.0 &                                    0.0 &                                    0.0 &                                    0.0 &  \cellcolor{MidnightBlue!40!white}11.00 &                                    0.0 &  \cellcolor{MidnightBlue!10!white}0.75 &   \cellcolor{MidnightBlue!20!white}8.5 &                                    0.0 &  \cellcolor{MidnightBlue!40!white}10.75 &                                   0.00 \\
\bottomrule
\end{tabular}}
\label{tab:main-results2}
\end{table*}

%% file: RQ3.tex
\begin{figure*}[!htbp]
    \centering
    \includegraphics[width=.95\linewidth]{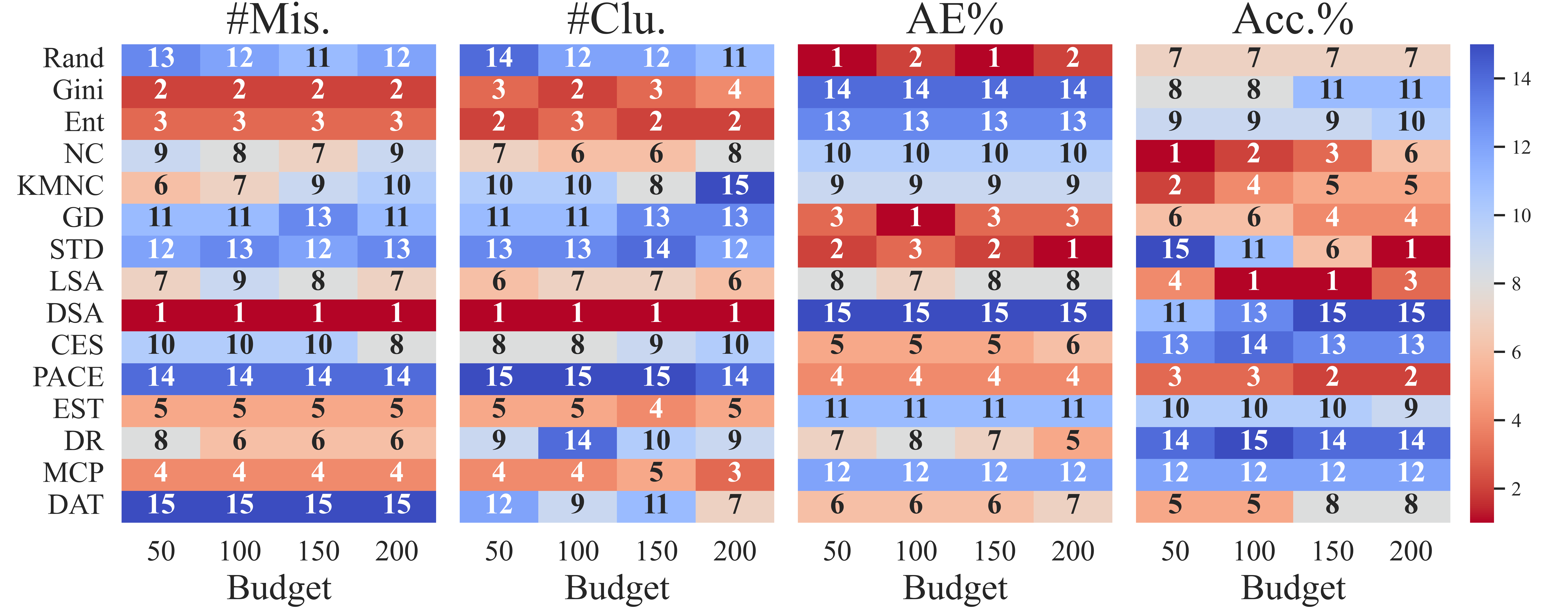}
    \caption{The rankings of each test selection metric across different budgets for each studied criterion (\#Mis., \#Clu., AE, Acc.\%). A rank of `1' indicates the best performance.}
    \label{fig:heatmap-rankings}
\end{figure*}

\subsection{RQ3. Performance of test selection methods under different selection budgets.}
 Figure \ref{fig:heatmap-rankings} shows the ranking of test selection metrics based on their average performance under different budgets for four criteria (\#Mis., \#Clu., AE\%, Acc.\%). The averages are calculated across all applicable experimental settings (e.g., 22 settings for classification-only metrics). A rank of `1' indicates the best performance. For instance, DSA's rank of `1' under \#Mis. at budget=50 means it selected the test suite that detects the highest number of mispredictions on average. This visualization clearly shows how each metric's performance changes with the budget. Under \textbf{fault detection} (\#Mis. and \#Clu.), most metrics' rankings are stable. For example, DSA consistently ranks first, followed by Gini and Ent. However, the ranking of KMNC decreases as the budget increases, likely because larger suites cover more neurons but not necessarily more faults. For \textbf{performance estimation} (AE\%), Rand consistently ranks first or second, while DSA consistently ranks last. Under \textbf{accuracy improvement} (Acc.\%), rankings are more volatile. STD's ranking improves from last to first as the budget increases, with its average accuracy improvement rising from 0.89\% to 4.76\%. A similar trend is observed in GD, which also selects the suite based on input diversity computed from features extracted in a black-box manner. Conversely, uncertainty-based (Gini, Ent) and coverage-based (NC, KMNC) metrics show a consistent decline in ranking.

\begin{tcolorbox}[colframe=red!40!black, sharp corners, leftrule={3pt}, rightrule={0pt}, toprule={0pt}, bottomrule={0pt}, left={2pt}, right={2pt}, top={3pt}, bottom={3pt}]
\textbf{Finding 7:} Metric performance for fault detection and performance estimation is largely stable across budgets, while performance for accuracy improvement shows considerable fluctuation.
\end{tcolorbox}

%% file: RQ4.tex
\subsection{RQ4. Time efficiency of test selection metrics.}
We evaluate the time efficiency of different test selection metrics by examining the total time costs (in seconds) of each metric when sampling 200 inputs (budget=200) from all settings for MNIST, Udacity, AndroZoo, and IMDb datasets. For example, for MNIST, we record the aggregated time costs of sampling 200 images from all 7 settings (i.e., ID 1, 2, 3, 14, 15, 16, 17). In this way, we can have a generalized overview of metrics' time efficiency under different settings. 
\begin{table*}[!htbp]
    \centering
\scriptsize
\addtolength{\tabcolsep}{-3.5pt}
\caption{The time costs (in seconds) of all 15 metrics when sampling 200 inputs from all settings under each dataset. Best values are highlighted.}
\begin{tabular}{c|ccccccccccccccc}
\toprule
\textbf{Dataset (settings)} & \textbf{Rand} & \textbf{Gini} & \textbf{Ent} & \textbf{NC} & \textbf{KMNC} & \textbf{GD} & \textbf{STD} & \textbf{LSA} & \textbf{DSA} & \textbf{CES} & \textbf{PACE} & \textbf{EST} & \textbf{DR} & \textbf{MCP} & \textbf{DAT} \\
\midrule
\textbf{MNIST} (7) & \cellcolor{gray!25}0.15 & 24.71 & 24.37 & 15.75 & 2561.05 & 101.38 & 98.82 & 447.92 & 974.57 & 288.33 & 19.38 & 1130.49 & 38.89 & 8.81 & 394.73\\
\textbf{Udacity} (8) & \cellcolor{gray!25}0.05 & - & - & 55.41 & 1960.33 & 325.62 & 320.57 & 83.07 & - & 760.74 & 32.11 & 137.23 & 82.72 & - & - \\
\textbf{AndroZoo} (7) & \cellcolor{gray!25}0.04 & 54.08 & 55.24 & 40.93 & 9574.37 & 59.50 & 54.62 & 219.13 & 311.30 & 743.47 & 43.80 & 915.01 & 95.47 & 22.62 & 272.73 \\
\textbf{IMDb} (8) & \cellcolor{gray!25}0.00 & 76.25 & 75.71 & 62.69 & 938.56 & 88.07 & 64.65 & 206.32 & 504.88 & 120.52 & 60.16 & 1492.51 & 266.59 & 35.71 & 430.56 \\
\midrule
\textbf{Average} & \cellcolor{gray!25}0.06 & 51.68 & 51.77 & 43.69 & 3758.58 & 143.64 & 134.67 & 239.11 & 596.92 & 478.26 & 38.87 & 918.81 & 120.92 & 22.38 & 366.01 \\
\bottomrule
\end{tabular}
\label{tab:time-efficiency}
\end{table*}

Table \ref{tab:time-efficiency} displays the time cost results. Rand (0.06s) is the fastest metric across all datasets. It is followed by clustering-based metrics (MCP: 22.38s, PACE: 38.87s) and uncertainty-based metrics (Gini: 51.68s, Ent: 51.77s). KMNC incurs the highest time cost on MNIST, Udacity, and Androzoo, whereas EST is the slowest on IMDb. The high time cost of KMNC arises from its iterative input selection process designed to maximize neuron coverage. EST's time cost varies because it relies on different auxiliary metrics (LSA for Udacity; DSA and confidence scores for others). This also explains the time difference between regression and classification datasets. Surprise-based metrics (LSA, DSA) are costly as they must compute and store activation traces for the entire training set to gauge `surprise'. For \textbf{fault detection}, a clear performance-speed trade-off exists: DSA is most effective (105.11 and 10.16 in \#Mis. and \#Clu., respectively) in classification tasks, while uncertainty metrics that rank second (Gini, Ent) are ten times faster but only half as effective, making them only suitable for time-critical applications. For \textbf{performance estimation}, the best-performing metrics (Rand, GD, STD) are also among the most time-efficient. For \textbf{retraining guidance}, while KMNC (4.37\%) and PACE (5.25\%) achieve the largest improvements for classification and regression, respectively, KMNC's high computational cost is a significant drawback. We recommend PACE for general use, as it offers a comparable 4.15\% improvement (a difference of only 0.22\%) for classification with far greater efficiency.

\begin{tcolorbox}[colframe=red!40!black, sharp corners, leftrule={3pt}, rightrule={0pt}, toprule={0pt}, bottomrule={0pt}, left={2pt}, right={2pt}, top={3pt}, bottom={3pt}]
\textbf{Finding 8:}  A clear performance-efficiency trade-off exists for best-performing metrics in fault detection. However, for performance estimation and retraining guidance, the most effective metrics are also highly efficient, offering no such trade-off.
\end{tcolorbox}

%% file: discussion.tex
\section{Discussion}
\subsection{Discussion of Findings}
We discuss our findings from three perspectives: practical implications for DNN testing, explanations of findings, and their relation to prior work.

\textbf{Practical Implications.}
Our results offer actionable guidance for selecting test metrics, retraining strategies, and fault identification pipelines under different objectives, data types, and budgets. (1) Testers should always prioritize inputs with high model surprise when aiming to detect faults in DL-enabled classifiers, irrespective of whether the test selection budget is tight or loose. In contrast, for DL-based regressors, test suites should be constructed by incorporating inputs that achieve broader neuron coverage. (2) Testers are encouraged to prioritize diverse inputs to achieve an accurate performance estimation, irrespective of whether the selection budget is tight or loose. (3) When the test selection budget is loose, select diverse inputs to retrain the model. This is based on the drastic improvement of STD's ranking as the budget increases, discovered in RQ3. (4) Type I retraining is recommended for sparse data (e.g., AndroZoo), whereas Type II is preferable for dense visual and textual data. (5) For efficient fault identification and targeted model repair, we recommend using (DeepDrebin, UMAP, DBSCAN) for malware data, (RoBERTa, UMAP, DBSCAN) for textual data, and (ResNet-50, UMAP, DBSCAN) for image data to conduct cluster-specific retraining for fault-targeted repair.

\textbf{Rationale Explanations of Findings.}
We provide rationales behind the main findings in each of the three testing objectives. Specifically, DSA is effective in fault detection since it captures boundary closeness by comparing activation distances to same-class versus different-class neighbors, where input close to the decision boundary is more likely to be misclassified \cite{kim2019guiding}. Moreover, EST shows a very inaccurate estimation in the classification problem. This may be due to its sampling-based strategy with DSA-based auxiliary variable, which over-selects mispredicted inputs, producing a suite that is not representative of the full test distribution. Diversity-based metrics better approximate the underlying data distribution, explaining their strong performance in estimation. Diversity-based metrics become increasingly effective for retraining guidance as the budget increases, as they progressively construct a more representative subset of the feature space.

\textbf{Alignment with Prior Findings.}
While some of our findings are consistent with prior work, they account for only a small portion of our results. For example, our observation that random selection can outperform specialized metrics for retraining under OOD settings aligns with \cite{hendrycks2016baseline}. Similarly, Devlin et al. \cite{devlin2019bert} report that uncertainty-based metrics outperform coverage-based metrics for fault detection in image classification, which we replicate and extend this finding across additional data modalities and OOD shifts. However, existing work \textit{lacks a unified evaluation of metrics designed for different objectives}. Our study addresses this gap and reveals that some metrics can outperform others even when they are not explicitly designed for the target objective.

\subsection{Threats to Validity}
The validity typology provides a system for classifying and improving inferences related to three validity types: internal validity, external validity, and construct validity \cite{anglin2024primer}.

\textbf{Internal Validity} concerns factors that could influence our results and the causal relationships we draw. To mitigate the threat in three aspects: (1) the implementation quality of the 15 test selection metrics, we mitigate the threats by closely following the original papers and using publicly available code. (2) hyperparameter manipulation, we mitigate this by using optimal values identified in the original papers for existing metrics (e.g., NC, KMNC), and for techniques used in the clustering pipeline, we employed established methods to determine the best hyperparameters. We provide details in our online repository for reproducibility. (3) Randomness with test input selection. We mitigate this by running experiments three times for metrics with randomness (EST, Rand, DAT, KMNC, NC, GD, STD, PACE, CES) and report the average value. We run only once for deterministic metrics (Gini, Ent, LSA, DSA, DR, MCP). 

\textbf{External Validity} concerns the generalizability of our findings beyond our experimental settings, which we address through three aspects: dataset and model selection, OOD scenarios, and selection budgets. To mitigate them, (1) we use four datasets with three modalities and 13 models across two tasks. (2) We construct five types of OOD scenarios (i.e., corruption, adversarial, temporal, natural, and label shifts). Specifically, adversarial and corrupted sets are constructed with multiple attack methods and severity levels to simulate a generalized real-world testing environment. We also select commonly used datasets from the literature to simulate natural covariate shifts and temporal shifts \cite{li2021can, hu2022empirical}. For label shifts, we randomly chose the label ratio, as the system under test may encounter arbitrary label distributions in reality. To further reduce the threat, we explore alternative label distributions with statistical analysis. We take MNIST with LeNet-5 and IMDb with Transformer as experiment subjects. Specifically, for MNIST, we simulate label shifts by sampling an opposite skewed distribution with 34\% for digit 9, 15\% each for digits 6-8, 5\% each for digits 3-5, 2\% each for digits 0-2. For IMDb, we sample the reviews with 20\% positive and 80\% negative. The results are shown in Table \ref{tab: external validity}, where the results are averaged over four budgets (50, 100, 150, 200). For MNIST (\textbf{MNIST-label2}), surprise-based and uncertainty-based metrics (DSA, Gini, Ent) achieve the best fault detection, yielding the highest \#Mis. and \#Clu. results. For performance estimation, EST, despite being designed for this objective, performs poorly, while diversity-encouraging metrics generally provide better estimates. Metrics tailored for retraining guidance (e.g., MCP, DAT) can be outperformed by others (e.g., STD for MNIST, Rand for IMDb). For IMDb (\textbf{IMDb-label2}), DSA again achieves the highest \#Mis., followed by Gini and Ent, which also lead in \#Clu. Diversity-based metrics achieve the top performance estimation, whereas EST remains ineffective. Overall, these results are consistent with our main findings, indicating that our conclusions remain robust under varied label distributions encountered in deployment.
\begin{table*}
\scriptsize
\addtolength{\tabcolsep}{-3.5pt}
\setlength\extrarowheight{-0.5pt}
\caption{Performance of 15 metrics under alternative label distributions (MNIST-label2 and IMDb-label2) and large selection budgets (IMDb-LB).}
\begin{tabular}{c|c|ccccccccccccccc}
\toprule
&  \textbf{Eval.} & \textbf{Rand} & \textbf{Gini} & \textbf{Ent} & \textbf{NC} & \textbf{KMNC} & \textbf{GD} & \textbf{STD} & \textbf{LSA} & \textbf{DSA} & \textbf{CES} & \textbf{PACE}  & \textbf{EST} & \textbf{DR} & \textbf{MCP} & \textbf{DAT} \\
\midrule
\multirow{4}{*}{\textbf{MNIST-label2}} & \#Mis.   &  2.25 & \cellcolor{MidnightBlue!40!white}59.0 & \cellcolor{MidnightBlue!40!white}60.75 & \textbf{5.25} & 4.0 & 3.0 & 2.25 & 1.0 & \cellcolor{MidnightBlue!40!white}62.25 & 2.5 & 3.0 & \cellcolor{MidnightBlue!10!white}14.5 & 0.75 & \cellcolor{MidnightBlue!20!white}41.25 & 2.5 \\

 &  \#Clu.   &  0.0 & \cellcolor{MidnightBlue!20!white}10.75 & \cellcolor{MidnightBlue!10!white}9.25 & 1.0 & 0.5 & 0.5 & 0.25 & 0.0 & \cellcolor{MidnightBlue!40!white}13.25 & 0.5 & 0.0 & \textbf{3.0} & 0.0 & \cellcolor{MidnightBlue!20!white}10.75 & 0.25 \\

 &  AE\%   &    \textbf{1.28} & 45.85 & 47.48 & 2.81 & 2.36 & \cellcolor{MidnightBlue!10!white}1.2 & \cellcolor{MidnightBlue!10!white}1.13 & \cellcolor{MidnightBlue!20!white}0.8 & 51.85 & \textbf{1.42} & \cellcolor{MidnightBlue!40!white}0.56 & 8.85 & \cellcolor{MidnightBlue!10!white}1.15 & 33.64 & \cellcolor{MidnightBlue!40!white}0.53 \\
 
 &  Acc.\%   &  0.93 & 0.61 & 0.34 & 0.87 & \textbf{1.05} & \cellcolor{MidnightBlue!10!white}1.18 & \cellcolor{MidnightBlue!40!white}1.3 & \textbf{1.07} & 0.78 & 0.31 & \cellcolor{MidnightBlue!20!white}1.21 & 0.67 & 0.07 & 0.6 & \cellcolor{MidnightBlue!20!white}1.21 \\

\midrule
\multirow{4}{*}{\textbf{IMDb-label2}} &  \#Mis.   &12.0 & \cellcolor{MidnightBlue!20!white}61.25 & \cellcolor{MidnightBlue!20!white}61.25 & 0.5 & \textbf{35.75} & 11.25 & 12.75 & 3.75 & \cellcolor{MidnightBlue!40!white}120.25 & 11.0 & 9.75 & \cellcolor{MidnightBlue!10!white}39.75 & \cellcolor{MidnightBlue!20!white}57.25 & \cellcolor{MidnightBlue!20!white}58.5 & 18.75 \\

 &   \#Clu.   & \textbf{2.25} & \cellcolor{MidnightBlue!20!white}3.25 & \textbf{2.25} & 0.0 & \cellcolor{MidnightBlue!40!white}4.0 & 1.5 & \textbf{2.0} & 0.75 & \cellcolor{MidnightBlue!20!white}3.25 & \cellcolor{MidnightBlue!10!white}2.75 & \textbf{2.0} & 0.5 & \cellcolor{MidnightBlue!20!white}3.5 & \cellcolor{MidnightBlue!10!white}2.75 & \cellcolor{MidnightBlue!20!white}3.25 \\

 &  AE\%   & \textbf{3.96} & 39.91 & 39.91 & 9.25 & 22.66 & \cellcolor{MidnightBlue!10!white}2.71 & \cellcolor{MidnightBlue!40!white}1.98 & 7.25 & 88.0 & \cellcolor{MidnightBlue!40!white}1.83 & \cellcolor{MidnightBlue!20!white}2.13 & 20.71 & 40.58 & 36.79 & \textbf{4.54} \\

 &   Acc.\%   & \cellcolor{MidnightBlue!40!white}1.28 & -0.02 & -1.27 & \textbf{0.66} & 0.1 & 0.11 & -0.49 & -3.68 & -3.07 & 0.32 & \cellcolor{MidnightBlue!20!white}0.9 & \cellcolor{MidnightBlue!10!white}0.82 & -1.61 & 0.22 & 0.1 \\
\midrule
\multirow{4}{*}{\textbf{IMDb-LB}} &  \#Mis.   &  109.33 & \cellcolor{MidnightBlue!20!white}356.17 & \cellcolor{MidnightBlue!20!white}356.17 & 2.0 & 96.33 & 114.33 & 117.0 & 53.5 & \cellcolor{MidnightBlue!40!white}610.17 & 113.83 & 91.17 & \cellcolor{MidnightBlue!10!white}284.5 & 148.17 & \cellcolor{MidnightBlue!20!white}358.67 & \textbf{193.33} \\

 &  \#Clu. & \textbf{2.83} & 2.33 & 2.0 & 0.0 & \cellcolor{MidnightBlue!20!white}3.83 & 2.5 & \cellcolor{MidnightBlue!10!white}3.0 & \cellcolor{MidnightBlue!20!white}3.67 & \textbf{2.67} & \textbf{2.83} & \cellcolor{MidnightBlue!20!white}3.5 & 2.17 & 2.33 & 2.17 & \cellcolor{MidnightBlue!40!white}16.67 \\
 
 &  AE\%   &   \cellcolor{MidnightBlue!10!white}1.29 & 32.11 & 32.11 & 15.23 & \textbf{2.77} & \cellcolor{MidnightBlue!40!white}0.44 & \cellcolor{MidnightBlue!20!white}0.68 & 8.41 & 66.37 & \cellcolor{MidnightBlue!10!white}1.26 & 3.23 & 22.28 & 4.57 & 32.48 & 10.36 \\
 
 &  Acc.\%   &   \cellcolor{MidnightBlue!20!white}4.5 & 3.42 & \textbf{3.63} & -0.1 & -0.18 & \cellcolor{MidnightBlue!40!white}4.9 & 1.43 & -1.03 & -3.43 & 0.37 & 2.9 & 2.23 & -0.53 & 0.89 & \cellcolor{MidnightBlue!10!white}3.9 \\
\bottomrule
\end{tabular}
\label{tab: external validity}
\end{table*}
 
(3) Finally, we evaluate budgets from 50 to 200 and observe that metric rankings are largely insensitive within this range. Since these budgets are relatively small, we further assess all 15 metrics under larger budgets using the IMDb–Transformer setting (ID 13), with selection sizes from 500 to 1,000, with an interval of 100. As shown in Table~\ref{tab: external validity} (\textbf{IMDb-LB}), the conclusions remain consistent: DSA and DAT achieve the best \#Mis. and \#Clu., respectively, diversity-based metrics (e.g., GD, STD) provide accurate performance estimation, and MCP and DAT do not yield top retraining performance.

\textbf{Construct Validity} concerns whether the study correctly identifies the operational measures for the concepts being investigated \cite{zhou2016map}. In our context, this means the selected evaluation criteria must reliably reflect the intended testing objectives. For fault detection, we employ the number of mispredictions (\#Mis.) and the number of clusters (\#Clu.) to evaluate the effectiveness. These two criteria are widely used in the literature \cite{feng2020deepgini, weiss2022simple, aghababaeyan2023black, attaoui2024supporting}. Specifically, \#Clu. alleviates the potential bias introduced by \#Mis. when mispredictions are attributed to the same fault. For retraining guidance and performance estimation, we assess the accuracy improvement (Acc.\%) and absolute error (AE\%), which are well-established and straightforward measurements \cite{chen2020practical, zhou2020cost, guerriero2024deepsample, guerriero2021operation, hu2022empirical, kim2019guiding, weiss2022simple}. Meanwhile, we also identify a more effective retraining process to accurately capture the improvement margin. 

%% file: Conclusion.tex
\section{Conclusion}
Our study presents an extensive study of 15 existing test selection metrics on benchmarking their performance on three testing objectives: fault detection, performance estimation, and retraining guidance. Through empirical analysis on five OOD types, four datasets with three modalities, and 13 DNNs, encompassing 1,640 unique experimental scenarios, we have derived critical insights for practitioners. For example, Metrics that encourage input diversity (Rand, GD, STD) are effective in performance estimation, outperforming those specifically designed for this objective. Meanwhile, metrics’ performance on fault detection and performance estimation is largely stable across budgets, while there is considerable fluctuation for retraining guidance. 

\section{Data Availability}
Our implementation is publicly available at \url{https://github.com/MetricsBenchmark/TestingBenchmark}.

\section*{Acknowledgments}
This work is supported by the National Natural Science Foundation of China under Grant 62502550, the Research Grants Council of Hong Kong (9229029), funds from CityU HK (9229192, 6000871), Shenzhen Science and Technology Program (KJZD20240903095700001), JST CRONOS Grant (No. JPMJCS24K8), and JSPS KAKENHI Grant (No.JP21H04877, No.JP23H03372, and No.JP24K02920), Canada CIFAR AI Chairs Program, the Natural Sciences and Engineering Research Council of Canada.